\documentclass[%
review,
twocolumn,
groupedaddress,
amsmath,amssymb,
prl,
]{revtex4-2}
\usepackage{graphicx}
\usepackage{hhline}
\usepackage{multirow}
\usepackage{color}
\usepackage{times}
\usepackage{url}
\usepackage{tikz}
\usepackage{tikz-feynman}
\usepackage{dsfont}
\usepackage{mathtools}
\usepackage{dcolumn}
\usepackage{setspace}
\usepackage{ulem}
\usepackage{physics}
\usepackage{comment}
\usepackage{hyperref}
\usepackage{bbm}
\usepackage[bottom]{footmisc}
\usepackage[toc,page,titletoc]{appendix}

\usepackage[most]{tcolorbox}
\usetikzlibrary{patterns}
\pgfdeclarepatternformonly{mystrikeout}{\pgfqpoint{-1pt}{-1pt}}{\pgfqpoint{11pt}{11pt}}{\pgfqpoint{10pt}{10pt}}%
{
  \pgfsetlinewidth{0.4pt}
  \pgfpathmoveto{\pgfqpoint{0pt}{0pt}}
  \pgfpathlineto{\pgfqpoint{10.1pt}{10.1pt}}
  \pgfusepath{stroke}
}
\newtcolorbox{tcbstrikeout}{breakable,
 enhanced jigsaw,
 opacityback=0,
 parbox=false,
 boxrule=0mm,
 top=0mm,bottom=0pt,left=0pt,right=0pt,
 boxsep=0pt,
 frame hidden,
 finish={\fill[pattern=mystrikeout] (frame.north west) rectangle (frame.south east);}
}

\begin{document}
	\title{Boundary obstructed topological superconductor in buckled honeycomb lattice under perpendicular electric field}
	
	\author{Rasoul Ghadimi}
	\author{Seung Hun Lee}
	\author{Bohm-Jung Yang}
	\email{bjyang@snu.ac.kr}
	
	\affiliation{Center for Correlated Electron Systems, Institute for Basic Science (IBS), Seoul 08826, Korea}
	\affiliation{Department of Physics and Astronomy, Seoul National University, Seoul 08826, Korea}
	\affiliation{Center for Theoretical Physics (CTP), Seoul National University, Seoul 08826, Korea}
	
	\date{\today}
\begin{abstract} 
In this work, we show that a buckled honeycomb lattice can host a boundary-obstructed topological superconductor (BOTS) in the presence of f-wave spin-triplet pairing (fSTP). 
The underlying buckled structure allows for the manipulation of both chemical potential and sublattice potential using a double gate setup.
Although a finite sublattice potential can stabilize the fSTP with a possible higher-order band topology, because it also breaks the relevant symmetry, the stability of the corner modes is not guaranteed. Here we show that the fSTP on the honeycomb lattice gives BOTS under nonzero sublattice potential, thus the corner modes can survive as long as the boundary is gapped.
 Also, by examining the large sublattice potential limit where the honeycomb lattice can be decomposed into two triangular lattices, we show that the boundary modes in the normal state are the quintessential ingredient leading to the BOTS.
 Thus the effective boundary Hamiltonian becomes nothing but the Hamiltonian for Kitaev chains, which eventually gives the corner modes of the BOTS.
	\end{abstract}
	\date{\today}
	 \maketitle

\begin{figure}
 	\centering
 	\includegraphics[width=1\linewidth]{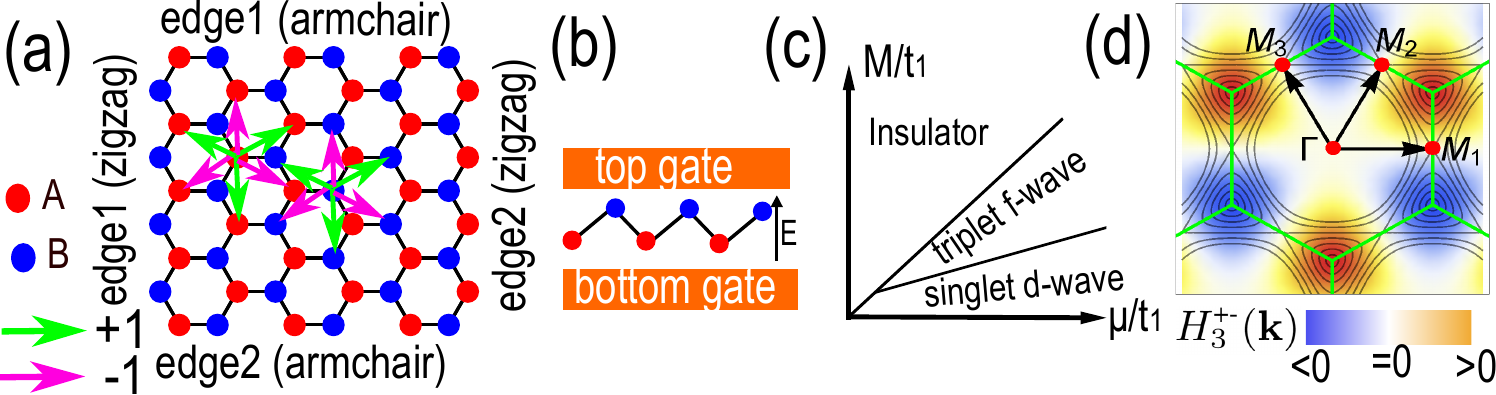}
 	\caption{
 			The fSTP in the honeycomb lattice.
 			(a) The top view of the honeycomb lattice with fSTP between next nearest-neighboring sites (shown with arrows).
 			(b) Double gate setup for tuning both $\mu$ and $M$ in a buckled honeycomb lattice. 
 			(c) The schematic superconducting phase diagram of a honeycomb lattice proposed in \cite{Zhang2015_f-wave_Electric_Induced_Doped_Silicene}.
 			(d) Momentum space distribution of a general fSTP ($H^{\text{+-}}_3(\mathbf{k})$ in Eq.~(\ref{BdGnewrep})).}
 	\label{fig:Fig1}
 \end{figure}

	
\textit{Introduction.---}
	Although superconductivity has not yet been observed in pristine graphene, it has been experimentally realized in related families  such as multi-layer~\cite{heikkila2022surprising,zhou2021superconductivity,Zhou2022BBGSC}, 
	twisted moir\'{e} bi-layer~\cite{Cao2018}, and alkali metal intercalated  graphene~\cite{yang2014superconducting,Kanetani_2012_Ca_intercalated_bilayer_graphene,Ichinokura_2016_SuperconductingCalcium_Intercalated_Bilayer_Graphene,Ludbrook2015SCLiDecoratedMonolayerGraphene,Toyama2022Ca-IntercalatedGrapheneSiC}. 
	Remarkably, underlying symmetries of the honeycomb lattice allow the emergence of various topologically nontrivial states, such as chiral p-wave and d-wave pairings~\cite{fukaya2016pairing,Wolf2022TSCHoneycombFermiSurface,Lee2019TSC_DiracHoneycomb,Ezawa2018ExactTSC_Hubbard_Fwave,Roy_Kekule_2010_Graphene,Zhou_2013GrapheneKekuleOrder,black2014chiral_d-wave,Black_Schaffer_2007_d_graphene,black2014chiral_d-wave_Mott,Gr_SC_Theory_s-wave_2007,Faye2015ppipGraphene,McChesney_2010_VHS_SC_GRAPHENE}. 
	Recently, it has been shown that the f-wave spin-triplet pairing (fSTP)~\cite{Song-Jin2021PRB_CompetingOrderHeavilyDopedHoneycom,Xianxin2019USC_jacutingaite_Pt2HgSe3,Lee2010NodalFwaveOpticalLAtticeHoneycomb,xiao2016possible_Singlet_Triplet_honeycomb,Crepel_PRB2022,Wolf_2022_Triplet_SC_Sn_Si111,Biderang_2022_TSC_SnSi,Honerkamp_2008_Honeycomb_f_did,Kiesel_2012_many-body-ins_SC_Graphene,Kagan2016_Long_Range_Coulomb_SC_Graphene,kagan2014kohnSCMonoAndBilayerGraphen,Nandkishore2014SC_Weak_hexagonal,chou2022acoustic,Chou_2021_Correlation_triplet_Graphene_moire,Fukaya2016SCDopedKaneMele,he2022superconductivity,JunctionAndScSymmetryTBG2022Guinea} can arise in the honeycomb lattice, deriving  higher-order topological superconductivity (HOTS)~\cite{li2021HOTPGraphene,scammell2021intrinsic} that hosts zero-energy Majorana corner modes, protected by both bulk gap and certain spatial symmetries,  such as inversion symmetry~\cite{WangPRB2021ConstructionHOTSHigherDimension,Trifunovic2021HOTPBandStructure,Schindler2018HOTI,Neupert2018,Yan2019PRLHOTSOddParitySC,Vu_2020_PRR_HOTS_TRI_DIII_C2,2020Kheirkhah_HOTS_Temperature-Driven,Ahn2020HOTSspin_polarized,Yan2019PRLHOTSOddParitySC,Wang2018WeakPairingHOTS,HsuPRL2018HOTSfromHOTI,Wang2018HighTemperatureHOTS,HsuInversionHOTSWTE2,Yan2018HOTSHighTempreture}.
	Also, fSTP was shown to be enhanced when there is a large sublattice potential~\cite{Crepel2021NewScMechanism,Crepel2022TripletSCdopedInsulator}.
 For instance, an electric field perpendicular to the plane of a low-buckled honeycomb lattice such as silicene can induce a considerable sublattice potential and stabilize fSTP (see Fig.~\ref{fig:Fig1}(b), and Fig.~\ref{fig:Fig1}(c))~\cite{Zhang2015_f-wave_Electric_Induced_Doped_Silicene}.
	However, as the sublattice potential breaks the essential symmetries that protect HOTS simultaneously, the stability of the corner modes is not guaranteed unless a distinct mechanism for their intrinsic protection exists.

	 In this letter, we show that  fSTP can host corner modes that remain intact even under considerable sublattice potential.
	Interestingly, the sublattice potential turns the HOTS into an extrinsic topological superconductor~\cite{WangPRB2021ConstructionHOTSHigherDimension,Trifunovic2021HOTPBandStructure,Wu2019HOTSSr2RuO4,Pan2019HOTSLatticeAssisted,Kheirkhah2020HOTSFLAT,Wu2020ZeemanFieldHOTS,Laubscher2019PRR_FTS_HOTS,Zhu20182DHOTSMagnetic,Wu2021HTC_HOTSFeSeTeMonolayer,ZhangPRR2020DetectionHOTS,Liu2018HOTS2DmagneticTI_HTC,Zhang2019HOTSandNodaSCFeSeTeHetro,VolpezPR2019LHOTSPiJunction,PhysRevB.106.085420,arxiv.2211.10905} or a boundary obstructed topological superconductor  (BOTS)~\cite{Ezawa_2020_Edge_corner_correspondence,Khalaf_Benalcazar_Hughes_Queiroz_2021_Boundary_obstructed,Tiwari2020ChiralDiracSC_BOTP,2020BOWuTP_SC_IronPnictides,Asaga2020BOTPDiracMagneticField,du2021acoustic,BOTSMagnetSuperconductorWong2022,PhysRevB.103.064512}, where corner modes are protected by the energy gap of the edge Hamiltonian, not the bulk Hamiltonian.
	We also propose a double gate setup that enables the realization of a tunable BOTS phase transition.

   In the following, we  first illustrate how unbroken symmetries can protect the BOTS phase, and then analyze the large sublattice potential limit of honeycomb lattices which effectively can be mapped into two triangular lattices.
    We demonstrate that the existence of boundary modes in the normal state (without superconductivity) is crucial for the BOTS (in the superconducting state).
    As fSTP acts effectively as a p-wave pairing for these boundary modes, the corner modes in the BOTS can be interpreted as the end modes of Kitaev chains \cite{Kitaev_2001} on the boundaries.

\textit{Model.---}
The  Bogoliubov-de-Gennes (BdG) Hamiltonian for the honeycomb lattice with $p_z$ orbital and fSTP is given by (see Sec.~S1 in the Supplemental Material (SM) for a derivation)~\footnote{
See Supplemental Material at **** for
a derivation of the BdG Hamiltonian, 
a derivation of the edge Hamiltonian and symmetry representation,
a derivation of the effective Hamiltonian in the large sublattice limit, 
a discussion on the spin-polarized case,
a discussion on BOTS in large spin-orbit coupling,
a discussion on the emergence of extrinsic HOTS by employing the valley hall effect,
numerical calculation of topological invariant for the slab of the honeycomb lattice, 
and study of the BOTS with different edge termination.} 
  \begin{align}\label{BdGnewrep}
 	&H_{\text{BdG}}(\mathbf{k})=t_1 H^{\text{++}}_1(\mathbf{k})\sigma_x\tau_z+t_1 H^{\text{-+}}_{1'}(\mathbf{k})\sigma_y\tau_z+ t_2 H^{\text{++}}_2(\mathbf{k})\tau_z\nonumber\\
  &-\mu \tau_z+M\sigma_z \tau_z+ \Delta_sH^{\text{+-}}_3(\mathbf{k}) s_x\tau_y,
 \end{align}
 where $t_1$, $t_2$, $\mu$, and $M$ are nearest-neighbor hopping, next-nearest-neighbor hopping, chemical potential, and sublattice potential, respectively ~\cite{Tsai2013GatedSilicene100spin-polarizedEelectrons,SiliceneEffectiveHamiltonian2011PRBLiu}. 
 We note that $\mu$ and $M$ can be changed by controlling the gate voltage (see Fig.~\ref{fig:Fig1}(b)).
In Eq.~(\ref{BdGnewrep}), $\sigma_{x,y,z}$, $s_{x,y,z}$, $\tau_{x,y,z}$ are the Pauli matrices for sublattice, spin, and electron-hole degrees of freedom.
The $H^{\eta_1,\eta_2}_n(k_x,k_y)$ are momentum dependent functions, in which $H^{\eta_1,\eta_2}_n(k_x,k_y){=}\eta_1 H^{\eta_1,\eta_2}_n(-k_x,k_y)$, $H^{\eta_1,\eta_2}_n(k_x,k_y){=}\eta_2 H^{\eta_1,\eta_2}_n(k_x,-k_y)$ and their explicit forms are provided in the SM~\cite{Note1}. 
In  Eq.~(\ref{BdGnewrep}), $\Delta_s{=}(\Delta_{A}{+}\Delta_{B})/2$ is the sublattice symmetric amplitude of fSTP, 
where $\Delta_{A}$ and $\Delta_{B}$ are the amplitude of fSTP on A and B sublattices, respectively. 
The $H^{\text{+-}}_3(\mathbf{k})$ in Eq.~(\ref{BdGnewrep}) is the Fourier transform of fSTP in the real space (see arrows inside Fig.~\ref{fig:Fig1}(a)), in which its zeros (pairing nodes) are located along $\Gamma-M_{1,2,3}$ (see Fig.~\ref{fig:Fig1}(d)).
In this letter, we are interested in a low doping limit in which the fSTP always gives  a fully gapped  superconductor to distinguish the in-gap corner modes. 
In the following, to investigate the topological state of Eq.~(\ref{BdGnewrep}),  we suppose a nonzero fSTP regardless of $\mu$ and $M$. 

\begin{figure}[t!]
	\centering
	\includegraphics[width=1\linewidth]{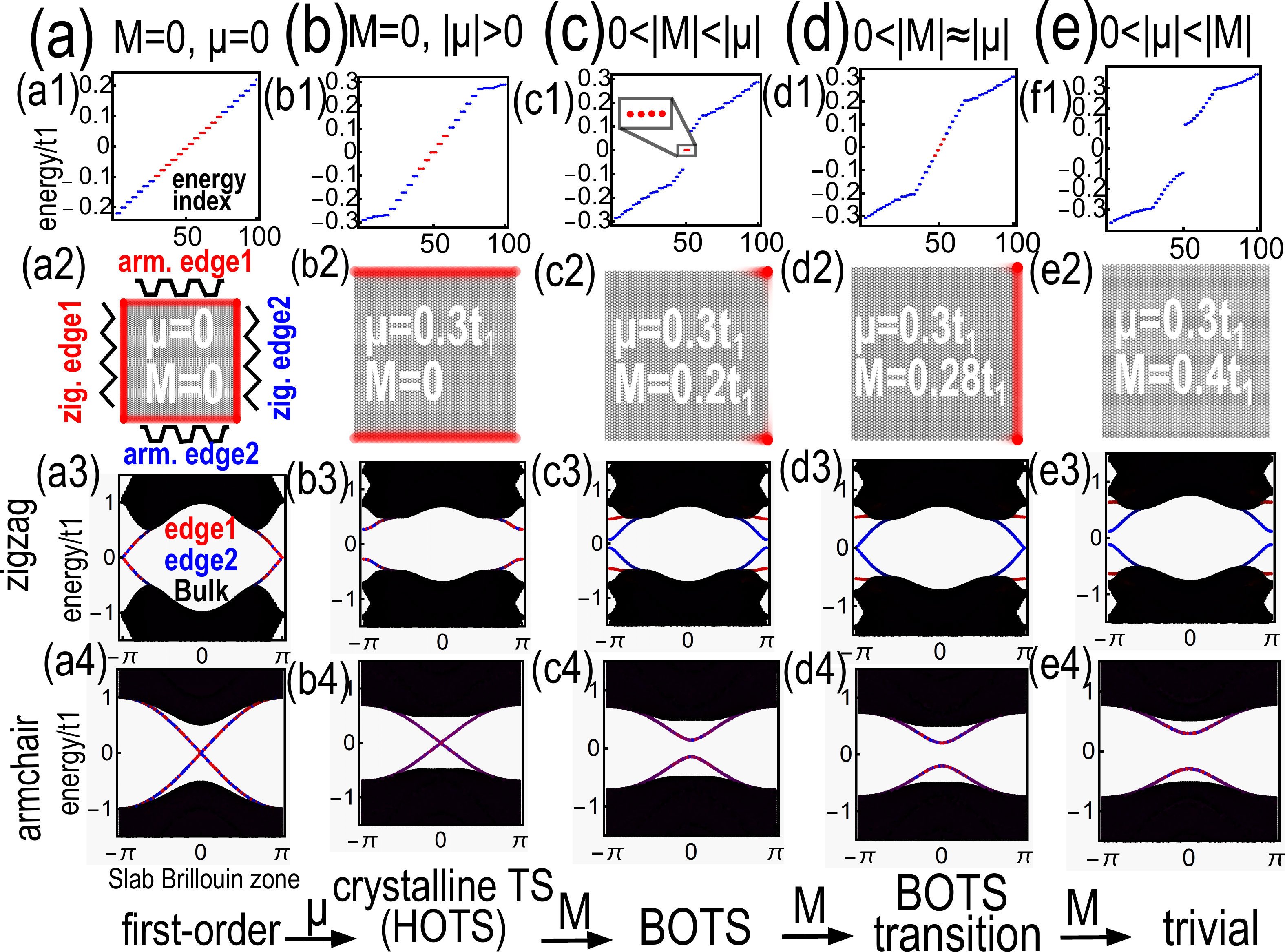}	
	\caption{ Sequence of topological phase transitions by varying $\mu$ and $M$.
	In the first row, we plot the energy spectrum for the square geometry  given in the second row. 
        In the second row, we plot the local density of states with red  dots for the energy states indicated by red in the first row. 
        In the third and fourth rows, we plot the energy spectrum for the zigzag and armchair strip geometries, respectively, with translational invariance along one direction where we show the probability of the wave function on the bulk and two edges (see Fig.~\ref{fig:Fig2}(a2)) by black and red/blue colors. 
        In these  figures, we set $t_2{=}0$,  $\Delta_{s}{=}0.1 t_1$, and $\mu$ and $M$  written inside the second row of each column.  
	}
	\label{fig:Fig2}
\end{figure}

\begin{figure}[t!]
	\centering
	\includegraphics[width=1\linewidth]{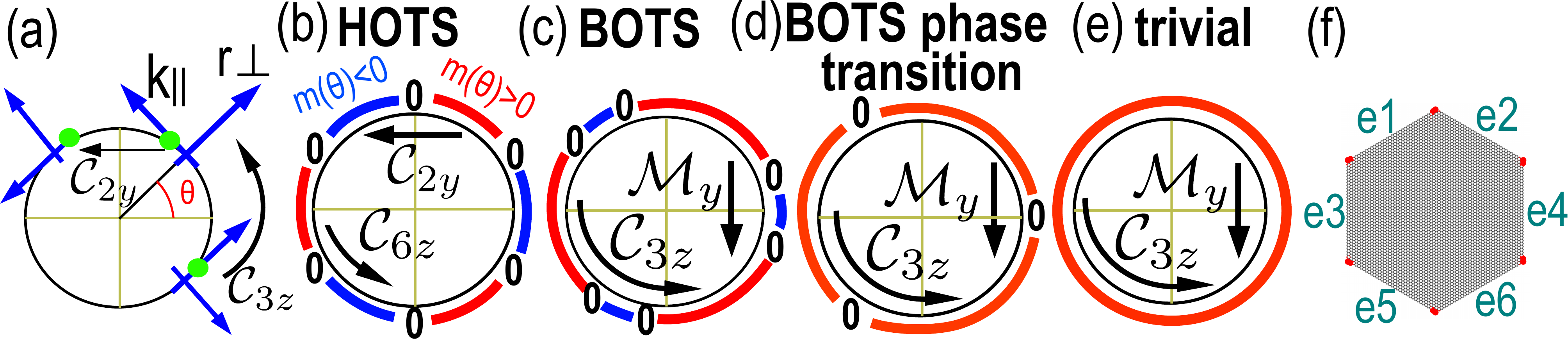}	
	\caption{(a) Disk geometry for edge theory.
(b)-(e) Schematic plot of the edge mass sign for the edge Hamiltonian in different topological phases. 
(b)	In  HOTS, certain symmetry operators such as $\mathcal{C}_{2y}$, $\mathcal{C}_{6z}$ force mass sign changing for different edges, and zero edge mass for symmetric edges.
(c) In BOTS, the remaining spatial symmetries such as $\mathcal{M}_y$, $\mathcal{C}_{3z}$  do not enforce mass sign reversal, but support a zero edge mass between edges (domains) with opposite mass signs.
(d) To eliminate these zero-edge masses, the remaining spatial symmetries enforce their pairwise annihilation at symmetric edges (such as the edge located on the right side).
(e) The trivial phase does not show any edge mass sign flip, and all boundaries are gapped out.  
(f) Hexagonal geometry with only zigzag edges in the HOTS or BOTS shows the same zero energy in-gap states localized at the corners (red spots).
	}
	\label{fig:Fig3}
\end{figure}

\textit{Results.---} 
We summarize the topological characteristic of Eq.~(\ref{BdGnewrep}) in the last row of Fig.~\ref{fig:Fig2}. 
In Fig.~\ref{fig:Fig2}(a), we show that when $\mu{=}M{=}0$ (we set $t_2{=}0$ for simplicity), the system hosts gapless states for all boundaries.
In Fig.~\ref{fig:Fig2}(b), we turn on $\mu{\ne}0$ and put $M{=}0$.
In this case, the gapless edge modes still exist along the armchair edges~(Fig.~\ref{fig:Fig2}(b4) and top and bottom edges of Fig.~\ref{fig:Fig2}(b2)).
Here, the system is a topological crystalline superconductor, where    $\mathcal{C}_{2y}{:}(x,z){\rightarrow}{-}(x,z)$ symmetric edges (armchair edges) remain gapless. 
In the absence of these symmetric edges, the system is HOTS (see Fig.~\ref{fig:Fig3}(f)), which has been already discussed in Ref.~\onlinecite{scammell2021intrinsic}. 
The HOTS arises from the fact that certain symmetry operators (e.g. inversion) flip the gap sign on the edge Hamiltonian, which gives in-gap states, called corner modes.
In Fig.~\ref{fig:Fig2}(c) we put $0{<}|M|{<}|\mu|$. 
In this case, although no gapless one-dimensional edge mode remains for the system, zero-energy corner modes still appear.
Note that depending on the signs of $M$ and $\mu$, the corner modes appear at different edges.
We will show that this phase is a BOTS.
The BOTS is protected by mirror $\mathcal{M}_y:y\rightarrow-y$ symmetry and corner modes survive as long as $\mathcal{M}_y$ symmetric edges (zigzag edges) remain gapped.
In Figs.~\ref{fig:Fig2}(d)-(e), we show that corner modes can be eliminated via a gap closing and reopening at one of the zigzag edges (right edge of Fig.~\ref{fig:Fig2}(d2) and edge-2 in Fig.~\ref{fig:Fig2}(d3)), which we call BOTS phase transition.
Note that the bulk remains gapped during this transition (black part of the energy spectrum in the third and fourth rows of Fig.~\ref{fig:Fig2}).

\textbf{Symmetries}
Equation~(\ref{BdGnewrep}) has particle-hole symmetry $\Xi{=}\tau_x\mathcal{K}{}_{[\mathbf k{\rightarrow}{-}\mathbf k]}$, and time-reversal symmetry  $\Theta{=}s_y\tau_z\mathcal{K}{}_{[\mathbf k{\rightarrow}{-}\mathbf k]}$, where $\mathcal{K}$ is the complex conjugate operator.
The system is also invariant under a $U(1)$ spin rotation, whose generator is given by $\mathcal{S}{=}s_z\tau_z$.
Moreover, Eq.~(\ref{BdGnewrep}) is  invariant under
a three-fold rotational symmetry $\mathcal{C}_{3z}{=}e^{i\tfrac{\pi}{3}s_z\tau_z}{}_{[\mathbf k{\rightarrow}\mathcal{R}_{2\pi/3}\mathbf k]}$,
and mirror symmetry $\mathcal{M}_y{=}s_y{}_{[{k_y\rightarrow}{-}k_y]}$,
where $\mathcal{R}_{\theta}$ rotates momentum vector $\mathbf k$ by $\theta$ around the out-of-plane direction. 
Furthermore, when $M{=}0$, Eq.~(\ref{BdGnewrep}) has inversion symmetry $\mathcal{P}{=}\tau_z\sigma_x{}_{[\mathbf k{\rightarrow}{-}\mathbf k]}$, 
a six-fold rotational symmetry $\mathcal{C}_{6z}{=}e^{i\tfrac{\pi}{6} s_z\tau_z}\sigma_x\tau_z{}_{[\mathbf k{\rightarrow}\mathcal{R}_{2\pi/6}\mathbf k]}$, 
and a two-fold rotational symmetries $\mathcal{C}_{2y}{=}s_y\sigma_x \tau_z{}_{[k_x{\rightarrow}{-}k_x]}$.
When $\mu{=}M{=}t_2{=}0$, Eq.~(\ref{BdGnewrep}) additionally has a local symmetry $\mathcal{L}{=}s_x\sigma_z\tau_y$ that commutes with $\mathcal{S}$. 
 
\textbf{First order topological phase} 
As we have shown in Fig.~\ref{fig:Fig2}(a), when $\mu{=}M{=}t_2{=}0$, the system supports gapless edge modes  at all edges, implying the existence of a first-order topological phase.
To see this, it is enough to rewrite  $H_{\text{BdG}}(\mathbf{k})$ using a unitary matrix $U$ (given in the SM \cite{Note1}) in the basis where both $U^\dagger\mathcal{S}U{=}s_z$ and $U^\dagger\mathcal{L}U{=}\tau_z$ (and $U^\dagger\sigma_zU{=}\sigma_z$) are diagonal and $ U^\dagger H_{\text{BdG}}(\mathbf{k})U{=}$
  \begin{align}\label{BdGnewrep2}
 t_1 H^{\text{++}}_1(\mathbf{k})\sigma_x{+}t_1 H^{\text{-+}}_{1'}(\mathbf{k})\sigma_y{+}H^{\text{+-}}_3(k)\Delta_s \tau_z  \sigma_z.
\end{align}
Therefore, in each sector of $s_z{=}\pm1$, $H_{\text{BdG}}$ decomposes to two copies of the celebrated Haldane model~\cite{Haldane_1988_HaldaneModel} with opposite Chern numbers; similar to the two-dimensional quantum spin Hall insulator model  proposed in graphene~\cite{Kane_Mele_2005_Z2QSHE,Kane_Mele_2005_QSHEGraphene}.

\textbf{Edge theory}
 Here, we derive an edge Hamiltonian at the boundary of a disk geometry to study the effect of nonzero $\mu$, $M$, and $t_2$ on the gapless boundary modes.
The edge Hamiltonian of Eq.~(\ref{BdGnewrep2}) composed of two counterpropagating chiral edge states,  can be written as 
\begin{equation}\label{SurfaceHamiltonian}
	H_{\text{b}}(\theta, k_{\parallel})= v_{\theta}   \tau_z  s_0 k_{\parallel},
\end{equation}
where $v_{\theta}$ is the Fermi velocity of the gapless edge modes that depends on the polar angle $\theta$, and $k_{\parallel}$ is the locally defined momentum tangent to the disk boundary (see Fig.~\ref{fig:Fig3}(a)).
Note that the edge Hamiltonian has to satisfy all symmetries of Eq.~(\ref{BdGnewrep2}),
whose representations can be obtained by considering their definitions regarding the edge Hamiltonian and commutation or anti-commutation relations between them (see Sec.~S2 of SM for a derivation \cite{Note1}).
The representations of  particle-hole, time-reversal, $\mathcal{L}$, and $\mathcal{S}$ symmetries are given by 
$\Xi{=}s_x\tau_z\mathcal{K}{}_{[k_{\parallel}{\rightarrow}{-}k_{\parallel}]}$,
$\Theta{=}is_y\tau_x\mathcal{K}{}_{[k_{\parallel}{\rightarrow}{-}k_{\parallel}]}$,
$\mathcal{L}{=}\tau_z$,
and $\mathcal{S}{=}s_z$, respectively.
For spatial symmetries, we 
find that there are two types of symmetry operations: type-1 symmetries (proportional to $\tau_y$ or $\tau_z$, such as $\mathcal{C}_{6z}$, $\mathcal{C}_{2y}$,  and $\mathcal{P}$), and type-2 symmetries  (proportional to $\tau_x$ or $\tau_0$, such as $\mathcal{C}_{3z}$,  and $\mathcal{M}_y$), where type-1 symmetries exchange the sublattice index, whereas type-2 symmetries preserve it.

\textbf{Possible edge mass}
We can study the topological feature of this system by understanding how  symmetry operations keep edge Hamiltonian gapless~\cite{Khalaf2018InversionHOTS}. 
The only edge mass that  anticommutes with Eq.~(\ref{SurfaceHamiltonian}), and respects both $\Xi$ and $\Theta$ is
$H_{\text{b.m}}(\theta)=m(\theta)\tau_x$.
However, since $\tau_x$ is odd under $\mathcal{L}$, edge Hamiltonian remains gapless ($m(\theta){=}0$) if $\mathcal{L}$ is present (i.e. $\mu{=}M{=}t_2{=}0$). 
On the other hand, spatial symmetries relate edge masses at different $\theta$. 
If we add a mass term that respects certain symmetries to the bulk BdG Hamiltonian,
we expect it gives an edge mass that also respects the same symmetries. 
The mass terms that are invariant under type-1 symmetries  give $m(\theta_{\text{p}}){=}{-}m(\theta)$, where $\theta_{\text{p}}$ means the symmetric partner of $\theta$ related by the given symmetry. 
Meanwhile, the  mass terms that are invariant under type-2 symmetries   give $m(\theta_{\text{p}}){=}{+}m(\theta)$.

\textbf{HOTS} 
Let us first assume that the system is invariant under type-1 symmetries, which force mass sign changing for symmetry related edges, giving zero energy corner modes~\cite{JackiwRebbi1976} (see Fig.~\ref{fig:Fig3}(b)).
 Then, the system belongs to HOTS~\cite{Khalaf2018InversionHOTS,2017_PRL_Langbehn_MirrorTopological_Ss_Ins}. 
We note that for the symmetric edges where $\theta_{\text{p}}{=}\theta$, such as $\theta{=}\tfrac{\pi}{2}$ (armchair edges) under $\mathcal{C}_{2y}$, the symmetry operation forces $m(\theta_{\text{p}}{\equiv}\theta){=}{-}m(\theta)=0$. 
Hence, the gapless boundary modes of the first-order topological phase are protected by symmetry operators on the symmetric edges. Accordingly, we can call this phase a $\mathcal{C}_{2y}$ protected topological crystalline superconductor~\cite{Zhang_Kane_Mele_2013_Topological_Mirror_Superconductivity}.
Both $\mu$ and $t_2$ are invariant under type-1 symmetries and keep the  gapless states  of  the armchair edges (Fig.~\ref{fig:Fig2}(b2) and Fig.~\ref{fig:Fig2}(b4)), while  gap out gapless states of  zigzag edges (Fig.~\ref{fig:Fig2}(b2) and Fig.~\ref{fig:Fig2}(b3)). Thus, the hexagonal geometry with only zigzag edges hosts only corner modes (see Fig.~\ref{fig:Fig3}(f)).

\textbf{BOTS} 
Now let's assume that the system is only invariant under type-2 symmetries, which do not force any topological gapless states, as symmetry-related edges obtain the same mass sign.
Despite this fact, the system can still have two different phases that cannot be smoothly connected without a gap closing at a symmetric edge (see Fig.~\ref{fig:Fig3}(c)-(e))~\cite{Ezawa_2020_Edge_corner_correspondence,Khalaf_Benalcazar_Hughes_Queiroz_2021_Boundary_obstructed}. 
For instance, $\mu$ and $M$ (and $t_2$) are invariant under type-2 symmetries.   
Therefore,  the HOTS (when $\mu{\ne}0$, $M{=}0$) can be smoothly transformed to BOTS by adding tiny $|M|$.
Accordingly, although small $M$ trivialize HOTS, still in-gap corner modes are protected because of the underlying BOTS phase (Fig.~\ref{fig:Fig2}(c2) or  Fig.~\ref{fig:Fig3}(f)).
These corner modes can be eliminated only by a BOTS phase transition through a gap closing at a symmetric edge (the zigzag edge located at the right side of Fig.~\ref{fig:Fig2}(d2) when $|M|{\approx}|\mu|$). 
 
\textbf{BOTS Phase transition} 
We can obtain a rough estimation of the BOTS transition for geometries with only zigzag edges. For instance, consider the hexagonal geometry given in Fig.~\ref{fig:Fig3}(f), where it has six edges that are labeled by ei (\{i=1,\dots,6\}) and their corresponding edge-masses are denoted by $m_{ei}$.
We first obtain the edge mass at one of the edges (for instance e4) and obtain others by symmetry considerations.  
To obtain $m_{e4}$, we need to find zero energy modes at that edge (when $\mu{=}t_2{=}M{=}0$) and then project $\mu$, $t_2$, and $M$  onto  these zero energy modes \cite{Winkler_Deshpande_2017_PRB_Effective_Hamiltonian_Grphene_M}, which leads to $m_{e4}{=}{-}\mu{-}2t_2{+}M$ (see Sec.~S3 in the SM~\cite{Note1}).
 The anti-commutation of a mass term (i.e. $M$)  with type-1 symmetries forces $m(\theta_{\text{p}}){=}m(\theta)$. 
 Thus, we obtain $m_{e3}{=}\mu{+}2t_2{+}M$, where signs of $\mu$ and $t_2$ are changed but $M$  keeps its sign. 
 Using type-2 symmetries, the e4, e1, and e5 (e2, e3, and e6) edges obtain the same edge mass.
As a result, nearby edges of Fig.~\ref{fig:Fig3}(f) obtain masses with opposite signs when $|M|<|\mu+2t_2|$ and give BOTS transition at $|M|=|\mu+2t_2|$, which leads to a trivial phase by further increasing $|M|$.

\begin{figure}[t!]
	\centering
	\includegraphics[width=1\linewidth]{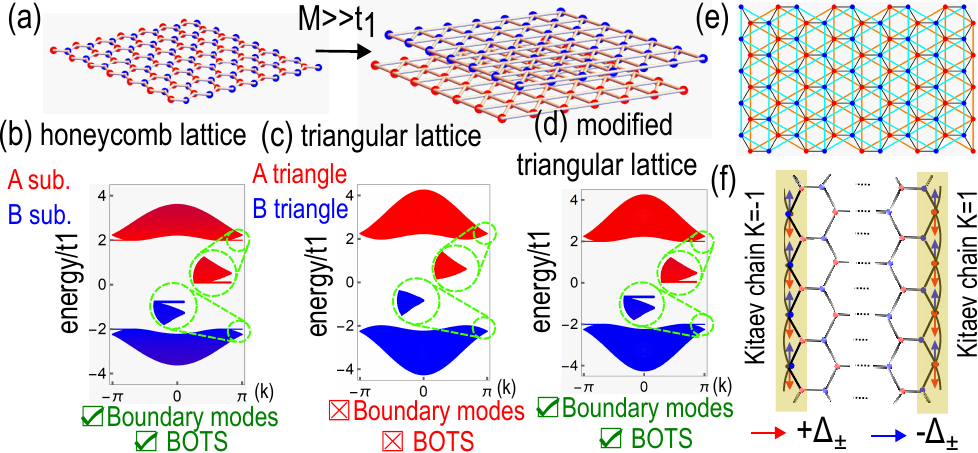}	
	\caption{
		(a) In the large sublattice potential $M$ limit, the honeycomb lattice is effectively decomposed into two triangular lattices. 
(b)  The normal state energy dispersion of the honeycomb lattice with strip geometry and zigzag edges, in which we only consider $t_1$ and $M$ (${=2t_1}$).
(c) The normal state energy dispersion of the effective triangular lattice (see the main text) with strip geometry and flat edges (see Fig.~\ref{fig:Fig4}(e)).
 (d)  Same plot as (c) obtained after modifying the onsite potential on the boundary vertices.
 The green insets in (b)-(d) magnify  energy  dispersion around $k{=}\pi$.
(e) Superimposed figure of honeycomb lattice with two effective triangular lattices. 
(f) Effective Kitaev chains labeled by $\kappa{=}{\pm}1$ reside at the zigzag edges of the  strip geometry of the honeycomb lattice.
	}
	\label{fig:Fig4}
\end{figure}

\textbf{Large sublattice potential limit}
 In the large $M$ limit, wave functions corresponding to the positive and negative energy bands of the honeycomb lattice are polarized on one sublattice (see Fig.~\ref{fig:Fig4}(b)).
Therefore in this regime, the honeycomb lattice can be effectively decomposed into two triangular lattices, which we denote by $\kappa{=}{\pm}1$~(see Fig.~\ref{fig:Fig4}(a)).
For instance, if we first take into account only $t_1$ and $M$, we can write an effective Hamiltonian on the two triangular lattices with nearest-neighbor hopping $t'{=}\tfrac{t_1^2}{2\kappa M}$ and onsite potential $V'{=}\kappa M{+}\tfrac{ z t_1^2}{2\kappa M}$ (see Sec.~S4 in SM~\cite{Note1}).
Here $z$ is the coordination number of a site that counts the number of the honeycomb links connected to it.
When $|M{|\gg}|t_1|$, we confirm that this effective Hamiltonian (by choosing $z$=3 for all sites) gives the identical bulk energy spectrum to the honeycomb lattice.
However, the zigzag boundary modes~\cite{PhysRevB.84.195452,PhysRevLett.89.077002} in the normal state of the honeycomb lattice are absent in the effective triangular lattices (compare Fig.~\ref{fig:Fig4}(b), and (c)).
As we have shown in Fig.~\ref{fig:Fig4}(d), we can restore the normal state boundary modes in the triangular lattices by modifying the effective onsite potential considering the fact that at the zigzag boundary vertices $z$=2 (see Fig.~\ref{fig:Fig4}(e)).
Interestingly, turning on fSTP in the triangular lattice does not lead to any corner modes unless we modify the onsite potential at boundary vertices.
Therefore, we can conclude that BOTS in the honeycomb lattice or modified triangular lattice are related to the presence of these normal state boundary modes. 

\textbf{Effective Kitaev-model and topological invariant}
Because normal state boundary modes are effectively localized at the  boundary vertices~\cite{CastroNeto2009}, we can write an effective 1D Hamiltonian~\cite{Ma2017ZigzagEdgetriplet,KitaevZigzagHoneycombpwaveRibeiroPRB2020}, in which the fSTP  acts as a p-wave pairing for these vertices (see Fig.~\ref{fig:Fig4}(f)).
Interestingly, this model can be considered as a generalization of the celebrated Kitaev model,  hosting Majorana end modes~\cite{Kitaev_2001,PollmannKitaev2017,PhysRevB.88.134523,Li2016}.
We can write a general effective Kitaev model including hopping, spin-orbit coupling (SOC),  $M$, and pairing  at the zigzag edges as 
\begin{equation}\label{Kitaev}
H_{\text{K}}(k){=}H_{\text{N}}(k)\tau_z{+}H_{\text{SO}}(k)s_z{+}H_{\text{R}}(k) s_x{+}H_{\Delta}(k)  s_x \tau_y,
\end{equation}
where we apply Fourier transformation using $k$, the momentum parallel to the chain.
In Eq.~(\ref{Kitaev}) $H_{\text{N}}(k)=-\mu+\kappa M+2t_2\cos(k)+F^{\kappa}(k)$, $H_{\text{SO}}(k)=2\kappa\lambda_{\text{SO}}\sin(k)$, $H_{\text{R}}(k)=2\lambda_{\text{R}}\sin(k)$, $H_{\Delta}(k)=2\Delta_{\kappa}\sin(k)$ and $F^{\kappa}(k)=\sum_{n=0}^{n_c} t_{n}^{\prime\kappa}\cos(n k)$, where $\lambda_{\text{SO}}$, $\lambda_{\text{R}}$, and $t_{n}^{\prime\kappa}$ are intrinsic SOC, Rashba SOC, and the effective hopping terms mediated by other (e.g. bulk) vertices, respectively.
For instance, it is possible to write $F^{\kappa}(k){\approx}\tfrac{(1{+}\cos(k))t_1^2}{\kappa M}$ by employing the effective Hamiltonian of the triangular lattice derived in the previous section as the simplest approximation.
The topological state of Eq.~(\ref{Kitaev}), is protected by both particle-hole $\Xi{=}\tau_x\mathcal{K}{}_{[k{\rightarrow}{-}k]} $, and time-reversal $\Theta{=}{s}_y\tau_z\mathcal{K}{}_{[k{\rightarrow}{-}k]}$ symmetries (see also Sec.~S5 and Sec.~S9 in the SM~\cite{Note1}),
where the system can have a nontrivial $Z_2$ topological phase and the Majorana end modes come in pairs~\cite{PhysRevB.88.134523}. 
Note that when $\lambda_R{=}\lambda_{\text{SO}}{=}0$, Eq.~(\ref{Kitaev}) is invariant under   $\Tilde{\mathcal{M}}_{y}=\tau_z {}_{[k{\rightarrow}{-}k]}$, which acts effectively as an inversion symmetry for the Kitaev chain.
Therefore, we can use the parity information $\xi(k)$ at $k{=}0,\pi$  to obtain $Z_2=\prod_{k=0,\pi}'\xi(k)$ invariant, where $'$ means the multiplication is done for only one state of the Kramer's pair~\cite{PhysRevB.88.134523}. 
Because at $k=0,\pi$, Eq.~(\ref{Kitaev}) reduces to $H_N(k)\tau_z$, we only need to check the sign changing between $H_N(0)$ and $H_{N}(\pi)$.
Therefore, we obtain $Z_2{=}\kappa\text{sgn}({-}\mu{+}\kappa M{-}2t_2)$, and the topological transition occurs at $|M|{=}|\mu{+}2t_2|$, consistent with the edge state analysis.
In the large $M$, normal state of two Kitaev chains $\kappa=\pm1$ are energetically separated  and accordingly for the geometry given in Fig.~\ref{fig:Fig4}(f), only one of $\kappa=\pm1$  can be topological at most.
Hence, we can interpret Fig.~\ref{fig:Fig2}(c2) containing a Kitaev chain with a topologically nontrivial phase on the  right zigzag edge.
The introduction of SOC does not break the nontrivial $Z_2$ topology of the effective Kitaev chain, as it preserves time-reversal symmetry. 
However, the presence of SOC can alter the equation defining the phase boundary, deviating slightly from $|M|=|\mu+2t_2|$, due to the emergence of other effective terms (see Sec.~S7 in the SM~\cite{Note1}).

Note that in Eq.~(\ref{Kitaev}), considering  sublattice dependent pairings $\Delta_{\kappa=\pm}{=}\Delta_{A, B}$, does not affect the topological properties of the effective Kitaev chains.
Therefore, the presence of nonzero asymmetric fSTP  $\Delta_a=\tfrac{\Delta_{A}{-}\Delta_{B}}{2}$ in the honeycomb lattice, resulting from an imbalance between the local density of states in the A and B sublattices under a nonzero sublattice potential, cannot alter the phase transition.
Furthermore, the dependence of the BOTS on the boundary modes in the normal states allows controlling the Majorana fermions using boundary engineering  (see Sec.~S10 in the SM~\cite{Note1}).
Additionally, a nontrivial topological state in the normal state always guarantees the existence of  boundary modes and can lead to Majorana corner modes in the presence of fSTP (see Sec.~S7 and Sec.~S8 of the  SM~\cite{Note1} for quantum spin hall effect and quantum valley hall effect \cite{Youngkuk_Valley_Silicene_2014}, respectively).

\textit{Conclusion.---}
In this letter, we have studied the topological features of the f-wave superconductor in buckled honeycomb structures with sublattice potential.
We first uncovered that despite breaking several symmetries by the sublattice potential, the remaining symmetries still protect BOTS with corner modes, and these corner modes disappear after a BOTS phase transition, mediated by gap closing on the symmetric edges.
Furthermore, we uncover that generating boundary modes in the normal state (using boundary engineering or manipulating gate voltages) is an important step to realizing BOTS in the presence of fSTP.
Our study can be extended to many recently discovered superconductors~\cite{heikkila2022surprising,zhou2021superconductivity,Zhou2022BBGSC,Cao2018} in which superconductivity may obtain an fSTP symmetry.

\begin{acknowledgements}
	R.G. thanks Hongchul Choi for the discussions.
R.G., S.H.L., and B.-J.Y. were supported by the Institute for Basic Science in Korea (Grant No. IBS-R009-D1),
Samsung  Science and Technology Foundation under Project Number SSTF-BA2002-06,
the National Research Foundation of Korea (NRF) grant funded by the Korea government (MSIT) (No.2021R1A2C4002773, and No. NRF-2021R1A5A1032996).
\end{acknowledgements}

\bibliography{refs}
\end{document}


	\title{Supplemental Material for Boundary obstructed topological superconductor in buckled honeycomb lattice under perpendicular electric field}
	
	\author{Rasoul Ghadimi}
		\author{Seunghun Lee }
	\author{Bohm-Jung Yang}
	
		\date{\today}
	\maketitle
 \tableofcontents
	\section{The derivation of the Bogoliubov-de-Gennes (BdG) Hamiltonian}
	\subsection{normal state}
	 The general Hamiltonian of a honeycomb material with $p_z$ orbital in the normal state  reads as,
\begin{align}\label{HR}
		&&H=-t_1\sum_{\expval{i,j},s}c^\dagger_{is}c_{js}+t_2\sum_{\expval{\expval{i,j}},s} c^\dagger_{is}c_{js} -\mu\sum_{i,s} c^\dagger_{is}c_{is}\nonumber\\
		&&+M\sum_{i,s} \kappa_i c^\dagger_{is}c_{is}+i\lambda_{SO}\sum_{\expval{\expval{i,j}},ss'} v_{ij}  c^\dagger_{is}[s_z]_{ss'}c_{js'}\nonumber\\
		&&+i\lambda_{R}\sum_{\expval{\expval{i,j}},ss'} c^\dagger_{is}[(\vec{s}\times\hat{d}_{ij})_{\hat{z}}]_{ss'}c_{js'},
\end{align}
where, $t_1$, $t_2$, $\mu$, $\lambda_{SO}$, $\lambda_{R}$, and $M$ are nearest-neighbor hopping, next-nearest-neighbor hopping, chemical potential, intrinsic spin-orbit coupling, Rashba spin-orbit coupling, and sublattice potential, respectively. 
In Eq.~(\ref{HR}), $\expval{i,j}$, and  $\expval{\expval{i,j}}$ indicate nearest-neighbor and next-nearest-neighbor links in the honeycomb structure.
The  $s_{\alpha=x,y,z}$ are usual Pauli matrices that act on the spin degree of freedom, and $ c^\dagger_{is}$, $ c_{is}$ are the creation and annihilation operator of an electron in $i$'th site with spin $s=\uparrow,\downarrow$, respectively.
We set $\kappa_{i\in \text{A}}=1$ and $\kappa_{i\in \text{B}}=-1$ where A and B denote the two sublattices of the honeycomb lattice (Fig.~1(a)).
In Eq.~(\ref{HR}), $\hat{d}_{ij}$ is the direction that connect $i$ to $j$ site and $v_{ij}=\tfrac{\vec{d}_{ik}\times\vec{d}_{kj}}{|\vec{d}_{ik}\times\vec{d}_{kj}|}$, where $k$ is the intermediate site that connect $i$ to site $j$ by a nearest-neighbor path.
By employing Fourier transformation $ c_{is}^\dagger=\frac{1}{\sqrt{N}}\sum_{\mathbf{k}\in B.Z} e^{i \mathbf{k}.r_i}  c_{\mathbf{k}\kappa_is}^\dagger$, $ c_{i s}=\frac{1}{\sqrt{N}}\sum_{\mathbf{k}\in B.Z} e^{-i \mathbf{k}.r_i}  c_{\mathbf{k}\kappa_is}$, we obtain
\begin{equation}\label{HND}
	H=\sum_\mathbf{k}\sum_{\kappa\kappa'=\{\text{A}, \text{B}\}}\sum_{s,s'=\{\uparrow,\downarrow\}}{c}^\dagger_{s\kappa\mathbf{k}} [H_\text{N}(\mathbf{k})]_{s\kappa,s'\kappa'} {c}_{s'\kappa'\mathbf{k}},
\end{equation}
where $\mathbf{k}=(k_x,k_y)$, and 
 \begin{align}\label{HNK}
	&H_\text{N}(\mathbf{k})=t_1 H^{\text{++}}_1(\mathbf{k})\sigma_x+t_1 H^{\text{-+}}_{1'}(\mathbf{k})\sigma_y+ t_2 H^{\text{++}}_2(\mathbf{k})-\mu \nonumber\\
	&+ \lambda_{SO}  H^{\text{+-}}_3(\mathbf{k})\mathbf{s}_z\sigma_z
	+ \lambda_{R} H^{\text{+-}}_4(\mathbf{k}) s_x+ \lambda_{R} H^{\text{-+}}_5(\mathbf{k}) s_y
	 +M\sigma_z.
\end{align}
In  Eq.~(\ref{HNK}), $\sigma_{\alpha=0,x,y,z}$ are the identity and Pauli matrices that act on the sublattice degree of freedom.
Furthermore, we suppress the direct product notation $\otimes$ between Pauli matrices, and $2\times 2$ identity matrices, and 
define
$H^{\text{++}}_1(\mathbf{k})=-2 \text{cos}\tfrac{k_x}{2} \text{cos} \tfrac{\sqrt{3} k_y}{2}-\text{cos} k_x$,
$H^{\text{-+}}_{1'}(\mathbf{k})=2 \text{sin}\tfrac{k_x}{2}  \text{cos} \tfrac{k_x}{2}-2 \text{sin}\tfrac{k_x}{2} \text{cos} \tfrac{\sqrt{3} k_y}{2}$,
$ H^{\text{++}}_2(\mathbf{k})=4 \text{cos} \tfrac{3 k_x}{2} \text{cos}  \tfrac{\sqrt{3} k_y}{2}+2\text{cos}  \sqrt{3} k_y$,
$ H^{\text{+-}}_3(\mathbf{k})= 2\text{sin}   \sqrt{3} k_y- 4\text{cos} \tfrac{3 k_x}{2}  \text{sin}   \tfrac{\sqrt{3} k_y}{2}$,
$ H^{\text{+-}}_4(\mathbf{k})=-2 \text{cos} \tfrac{3 k_x}{2} \text{sin} \tfrac{\sqrt{3} k_y}{2}-2\text{sin} \sqrt{3} k_y$, and
$ H^{\text{-+}}_5(\mathbf{k})=2 \sqrt{3}  \text{sin} \frac{3 k_x}{2} \text{cos} \frac{\sqrt{3} k_y}{2}$, respectively.

\begin{figure}
 	\centering
 	\includegraphics[width=1\linewidth]{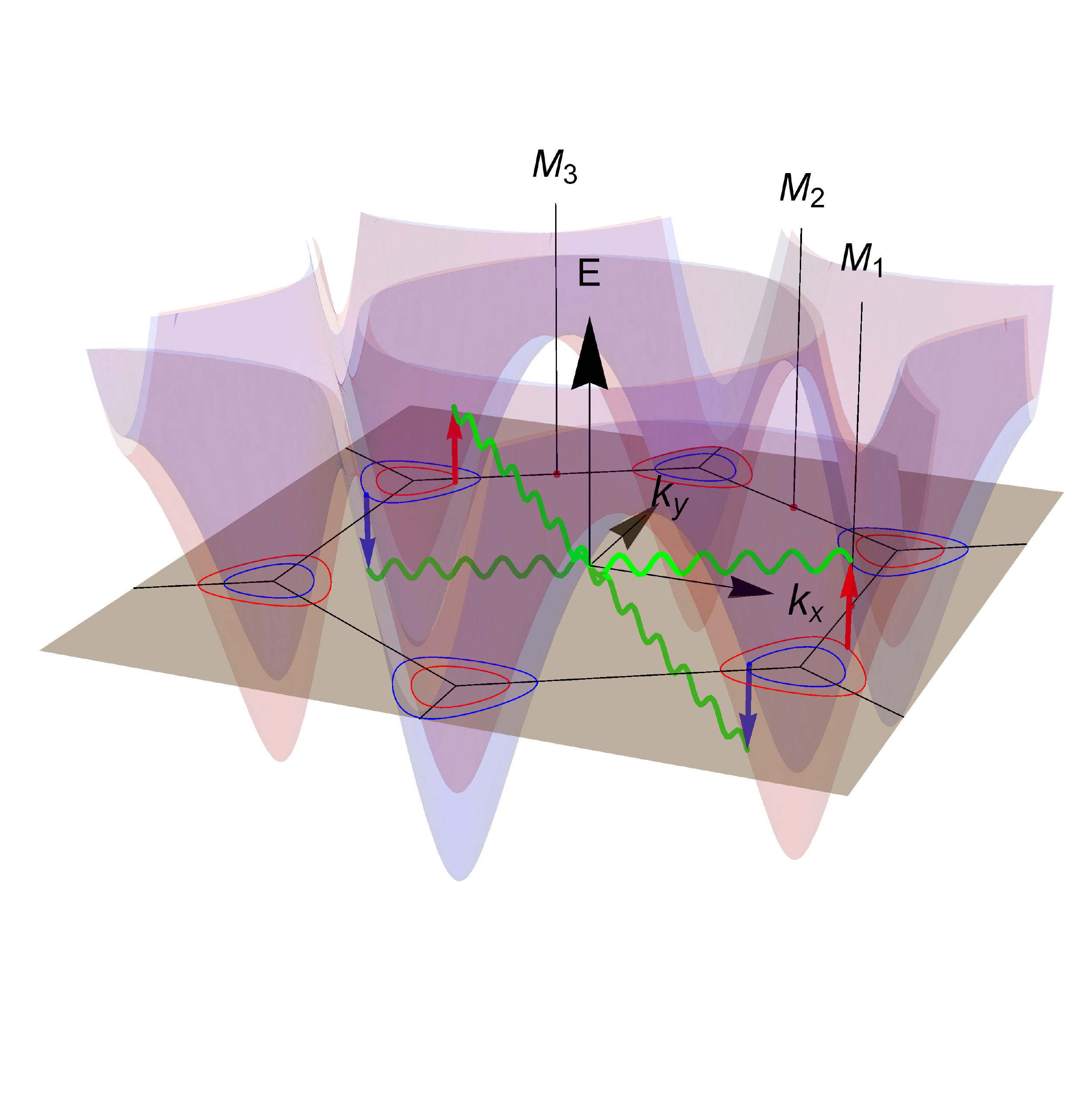}
 	\caption{
 			Energy dispersion (normal state) of the honeycomb lattice for positive energy, when $M\ne0$ and $\lambda_{\text{SO}}\ne0$. The intersection of energy dispersion (with some $\mu$) is shown with blue and red contours, indicating their spin characteristic. The wiggly green lines show the instability toward non-FFLO superconductivity between opposite spins.  }
 	\label{fig:dispersion}
 \end{figure}
\subsection{pairing potential}
The  energy dispersion of the Eq.~(\ref{HNK}) in the presence of both $\mu$ and $\lambda_{\text{SO}}$ but without $\lambda_{\text{R}}=0$ is shown in Fig.~\ref{fig:dispersion}.
As we can see, the system has two valleys, and the Fermi contours along them are spin-polarized, and its polarization flips in different valleys.
Considering the Fermi-contour geometry (red and blue contours in Fig.~\ref{fig:dispersion}), and in the absence of Fulde–Ferrell–Larkin–Ovchinnikov (FFLO) superconductivity, the pairing can occur among electrons with opposite spin and momenta which can lead to both singlet and triplet pairing instabilities. 
Furthermore, in the buckled honeycomb structure, such as silicene, we can tune both $\mu$ and $M$ using a double-gate setup (see Fig.~1(b)), which may enhance or change superconductivity instability.
Indeed it is shown that the triplet f-wave pairing can be dominated over a singlet d-wave channel in the presence of large $M$ by polarizing wave function in one of the sublattices \cite{Zhang2015_f-wave_Electric_Induced_Doped_Silicene} (see Fig.~1(c)).  
The f-wave pairing has $\text{sin} 3\theta_{ij}$ characteristic, where  $\theta_{ij}$ is the azimuthal angle from the horizontal direction.  
Therefore, the pairing potential between the nearest-neighbor vertices is zero, while for the next-nearest-neighbors, the sign of pairing flips six times by rotating around a given site (see Fig.~1(a)). 
In the main letter, we assume that $\lambda_R,\lambda_{SO}\ll M$  which can be satisfied for low buckled honeycomb structure under a strong electric field, and we expect f-wave pairing can be dominated even in the presence of small spin-orbit coupling. 
Because of $M\ne0$, the f-wave pairing potential $\Delta_{A}$, $\Delta_{B}$ can be different for A and B sublattices, respectively. 
The f-wave pairing in real space can be modeled by
\begin{equation}\label{HSC}
H_{sc}=\sum_{\expval{\expval{i,j}}} \sum_{ss'} v_{ij} (\kappa_i \Delta_s+\Delta_a) c^\dagger_{is} [s_x]_{ss'} c^\dagger_{is'},
\end{equation}
where we define symmetric $\Delta_s=(\Delta_{A}+\Delta_{B})/2$, and asymmetric $\Delta_a=(\Delta_{A}-\Delta_{B})/2$ pairing potential. 
Note that $v_{ij}=\kappa_i \text{sin}3\theta_{ij}$.
We can multiply Eq.~(\ref{HSC}) by any arbitrary phase factor, which although may change the form of the  Hamiltonian and symmetries representation, does not have any physical importance. 

\subsection{BdG Hamiltonian}
By obtaining Fourier transformation of Eq.~(\ref{HSC}), 
$H_{sc}=-i\sum c^\dagger_{s\kappa\mathbf{k}} H^{\text{+-}}_{3}(\mathbf{k})(k)[(\Delta_s+\sigma_z \Delta_a)s_x ]_{s\alpha,s'\alpha'} c^\dagger_{s'\kappa\mathbf{k}}$.
We can write the Bogoliubov–de Gennes (BdG)   
Hamiltonian as
\begin{equation}\label{BdG}
	H_{\text{BdG}}= \sum a^\dagger_{s\kappa\tau\mathbf{k}} [H_{\text{BdG}}(\mathbf{k})]_{s\kappa\tau,s'\kappa'\tau'} a_{s'\kappa'\tau'\mathbf{k}},
\end{equation}
where $\tau=\pm 1$ indicate electron and hole in the Nambu space;  $a^\dagger_{s\kappa(+1)\mathbf{k}}=c^\dagger_{s\kappa\mathbf{k}}$, $a^\dagger_{s\kappa(-1)\mathbf{k}}=c_{s\kappa-\mathbf{k}}$,
$a_{s\kappa(+1)\mathbf{k}}=c_{s\kappa\mathbf{k}}$, $a_{s\kappa(-1)\mathbf{k}}=c^\dagger_{s\kappa-\mathbf{k}}$, and we assume summation over repeated indices. 
The $H_{\text{BdG}}(\mathbf{k})$ is given as follow
  \begin{align}\label{BdGnewrep}
 	&H_{\text{BdG}}(\mathbf{k})=t_1 H^{\text{++}}_1(\mathbf{k})\sigma_x\tau_z+t_1 H^{\text{-+}}_{1'}(\mathbf{k})\sigma_y\tau_z+ t_2 H^{\text{++}}_2(\mathbf{k})\tau_z\nonumber\\
 	&+ \lambda_{SO}  H^{\text{+-}}_3(\mathbf{k})s_z\sigma_z 
 	+ \lambda_{R} H^{\text{+-}}_4(\mathbf{k}) s_x+ \lambda_{R} H^{\text{-+}}_5(\mathbf{k}) s_y \tau_z
 	\nonumber\\
 	&-\mu \tau_z+M\sigma_z \tau_z+ H^{\text{+-}}_3(\mathbf{k})(\Delta_s+\sigma_z \Delta_a) s_x\tau_y,
 \end{align}
where $\tau_{0,x,y,z}$ are identity and the Pauli matrices that act on the electron-hole degrees of freedom. 

\subsection{Symmetries}
In this section, we present the extended symmetries of  (\ref{BdGnewrep}). For simplicity lets first assume that honeycomb lattice is unbuckled.
Equation~(\ref{BdGnewrep}) generally (in the existence of all terms) has particle-hole symmetry $\Xi=\tau_x\mathcal{K}[\mathbf k \rightarrow -\mathbf k]$,  time-reversal symmetry  $\Theta=s_y\tau_z\mathcal{K}[\mathbf k \rightarrow -\mathbf k]$, and chiral symmetry  $\Pi=s_y\tau_y$, where $\mathcal{K}$ is the complex conjugate operator.  

Moreover, Eq.~(\ref{BdGnewrep}) is  invariant under
a three-fold rotational Symmetry 
$\mathcal{C}_{3z}=e^{i\tfrac{\pi}{3} s_z\tau_z}[\mathbf k \rightarrow \mathcal{R}_{2\pi/3}\mathbf k]$,
mirror symmetry $\mathcal{M}_y=s_y [{(k_x,k_y)} \rightarrow {(k_x,-k_y)}]$,
and a  two-fold rotational Symmetry 
$\mathcal{C}_{2x}= s_x \tau_z [{(k_x,k_y)} \rightarrow {(k_x,-k_y)}]$ ($\lambda_R=0$),
in addition to $U(1)$ spin rotation around the out-of-plane direction, whose generator is given by
$\mathcal{S}=s_z\tau_z$ ($\lambda_R=0$).

Furthermore, when $\Delta_a=M=0$, Eq.~(\ref{BdGnewrep}) 
has inversion symmetry 
$\mathcal{P}=\tau_z\sigma_x [\mathbf k \rightarrow -\mathbf k]$ ($\lambda_R=0$),
a mirror symmetry 
$\mathcal{M}_x=s_x \sigma_x [{(k_x,k_y)} \rightarrow {(-k_x,k_y)}]$,
and a six-fold rotational Symmetry 
$\mathcal{C}_{6z}=e^{i\tfrac{\pi}{6} s_z\tau_z}\sigma_x\tau_z[\mathbf k \rightarrow \mathcal{R}_{2\pi/6}\mathbf k]$ 
and two two-fold rotation symmetries 
$\mathcal{C}_{2z}=s_z\sigma_x [\mathbf k \rightarrow -\mathbf k]$
and 
$\mathcal{C}_{2y}=s_y\sigma_x \tau_z[{(k_x,k_y)} \rightarrow {(-k_x,k_y)}]$ ($\lambda_R=0$).

When $\mu{=}M{=}t_2{=}0$, Eq.~(\ref{BdGnewrep}) has an additional local symmetry $\mathcal{L}=s_x\sigma_z\tau_y$ ($\lambda_{so}=0$) that commutes with $\mathcal{S}$. 

On the other hand, when $\Delta_a=\Delta_s=0$, Eq.~(\ref{BdGnewrep}) has two additional symmetries regarding $U(1)$ gauge symmetry $\mathcal{T}=\tau_z$ and spin rotational Symmetry $\Tilde{\mathcal{S}}=s_z$   ($\lambda_R=0$), where $\mathcal{S}=\Tilde{\mathcal{S}}\mathcal{T}$. 

Note that considering buckling property, some symmetry operators such as $\mathcal{M}_x$, and $\mathcal{C}_{2x}$ are not true spatial symmetry of the system. However, they are related to the other true spatial symmetry of the system by multiplication of $\mathcal{S}$, for instance, $\mathcal{M}_x=-i\mathcal{C}_{2y}\mathcal{S}$, and  $\mathcal{C}_{2x}=i\mathcal{M}_y\mathcal{S}$, respectively. 

\subsection{Basis transformation}
In studying the first order topological phase and deriving the edge Hamiltonian we rewrite Eq.~(\ref{BdGnewrep}) in a basis that $\mathcal{L}$,  $\mathcal{S}$, and $\sigma_z$ are diagonal. 
we use 
\begin{equation}
    U=e^{i\pi s_x \sigma_z  (-\tfrac{1}{4} +\tfrac{1}{3\sqrt{3}} \tau_x+\tfrac{1}{3\sqrt{3}} \tau_z)+i\pi\tfrac{1}{3\sqrt{3}} \tau_y}
\end{equation}
and get ($\lambda_{SO}=\lambda_R=\Delta_a=0$) 
  \begin{align}\label{BdGnewrep2}
U^\dagger H_{\text{BdG}}(\mathbf{k})U=
& t_1 H^{\text{++}}_1(\mathbf{k})\sigma_x+t_1 H^{\text{-+}}_{1'}(\mathbf{k})\sigma_y+\tau_z H^{\text{+-}}_3(k)\Delta_s \sigma_z+ \nonumber\\
&(t_2 H^{\text{++}}_2(k)-\mu )\tau_x+M\sigma_z\tau_x.
\end{align}

\begin{table*}
\centering
\begin{tabular}{|c| c|c| c| c c| c c c c c c c c c c||c|c|c|c|c|} 
\hline
&\multicolumn{2}{|c|}{bulk representation} &\multicolumn{13}{|c|}{algebra between symmetry operators: $XX'- s X'X=0$} & \multicolumn{3}{|c|}{edge representation}& \multicolumn{2}{|c|}{algebra} \\
 \hline
 $X$ & operation & $\mathbf{k}\rightarrow$ &$X^2$& $\Xi$ & $\Theta$&$\mathcal{L}$&$\mathcal{S}$&$\mathcal{M}_y$&$\mathcal{C}_{2x}$&$\mathcal{P}$&$\mathcal{M}_x$&$\mathcal{C}_{2z}$&$\mathcal{C}_{2y}$&$\mathcal{C}_{3z}$&$\mathcal{C}_{6z}$
 & operation & $\theta\rightarrow$& $k_{\parallel}\rightarrow$&$\tau_z$& $\tau_x$ \\
 \hline\hline
$\Xi$ & $is_x\tau_z\sigma_z \mathcal{K}$&$-\mathbf{k}$ 
&$+1$& $+1$ & $+1$&$+1$&$-1$&$-1$&$-1$&$-1$&$+1$&$+1$&$+1$&$+1$&$-1$&$s_x\tau_z\mathcal{K} $&$\theta$&$-k_{\parallel}$&$+1$& $-1$ \\
$\Theta$ &$ s_y\tau_x \mathcal{K} $&$-\mathbf{k}$  &$-1$&
$+1$ & $+1$&$-1$&$-1$&$-1$&$-1$&$+1$&$-1$&$-1$&$-1$&$+1$&$+1$&$is_y\tau_x\mathcal{K} $&$\theta$&$-k_{\parallel}$&$-1$& $+1$ \\
$\mathcal{L}$ & $\tau_z$&$\mathbf{k}$  &$+1$& $+1$ & $-1$&$+1$&$+1$&$-1$&$-1$&$+1$&$-1$&$+1$&$-1$&$+1$&$+1$&$\tau_z$&$\theta$&$k_{\parallel}$&$+1$& $-1$ \\
$\mathcal{S}$ & $s_z$&$\mathbf{k}$  &$+1$& $-1$ & $-1$&$+1$&$+1$&$-1$&$-1$&$+1$&$-1$&$+1$&$-1$&$+1$&$+1$&$s_z$&$\theta$&$k_{\parallel}$&$+1$& $+1$ \\
$\mathcal{M}_y$ & $s_y \tau_x $&$(k_x,-k_y) $  &$+1$& $-1$ & $-1$&$-1$&$-1$&$+1$&$-1$&$+1$&$-1$&$-1$&$+1$&$\dots$&$\dots$&$s_y\tau_x$&$-\theta$&$-k_{\parallel}$&$-1$& $+1$ \\
$\mathcal{C}_{2x}$ & $ s_x\tau_x $&$ (k_x,-k_y)$  &$+1$& $-1$ & $-1$&$-1$&$-1$&$-1$&$+1$&$+1$&$+1$&$-1$&$-1$&$\dots$&$\dots$&$s_x\tau_x$&$-\theta$&$-k_{\parallel}$&$-1$& $+1$ \\
$\mathcal{P}$ & $ \sigma_x $&$- \mathbf{k} $  &$+1$& $-1$ & $+1$&$+1$&$+1$&$+1$&$+1$&$+1$&$+1$&$+1$&$+1$&$+1$&$+1$&$s_z\tau_z$&$\theta+\pi$&$k_{\parallel}$&$+1$& $-1$ \\
$\mathcal{M}_x$ & $s_x \tau_x \sigma_x $&$(-k_x,k_y) $  &$+1$& $+1$ & $-1$&$-1$&$-1$&$-1$&$+1$&$+1$&$+1$&$-1$&$-1$&$\dots$&$\dots$&$s_y\tau_y$&$\pi-\theta$&$-k_{\parallel}$&$-1$& $-1$ \\
$\mathcal{C}_{2z}$ & $s_z \sigma_x $&$- \mathbf{k} $  &$+1$& $+1$ & $-1$&$+1$&$+1$&$-1$&$-1$&$+1$&$-1$&$+1$&$-1$&$+1$&$+1$&$\tau_z$&$\theta+\pi$&$k_{\parallel}$&$+1$& $-1$  \\
$\mathcal{C}_{2y}$ & $s_y \tau_x \sigma_x $&$ (-k_x,k_y) $  &$+1$& $+1$ & $-1$&$-1$&$-1$&$+1$&$-1$&$+1$&$-1$&$-1$&$+1$&$\dots$&$\dots$&$s_x\tau_y$&$\pi-\theta$&$-k_{\parallel}$&$-1$& $-1$ \\
$\mathcal{C}_{3z}$& $\tfrac{\mathbbm{1}+i \sqrt{3}s_z}{2}$&$  \mathcal{R}_{2\pi/3}\mathbf{k} $  &$-\mathcal{C}^{-1}_{3z}$& $+1$ & $+1$&$+1$&$+1$&$\dots$&$\dots$&$+1$&$\dots$&$+1$&$\dots$&$+1$&$+1$&$\tfrac{\mathbbm{1}+i \sqrt{3}s_z}{2}$&$\theta+2\pi/3$&$k_{\parallel}$&$+1$& $+1$ \\
$\mathcal{C}_{6z}$ & $\tfrac{\sqrt{3}\sigma_x+i s_z\sigma_x }{2}$&$ \mathcal{R}_{2\pi/6}\mathbf{k} $  &$\mathcal{C}_{3z}$& $-1$ & $+1$&$+1$&$+1$&$\dots$&$\dots$&$+1$&$\dots$&$+1$&$\dots$&$+1$&$+1$&$\tfrac{\sqrt{3}s_z \tau_z+i \tau_z}{2}$&$\theta+2\pi/6$&$k_{\parallel}$&$+1$& $-1$ \\
 \hline
\end{tabular}
\caption{Symmetry operators (in the bulk representation Eq.~(\ref{BdGnewrep2}) and edge representation Eq.~(\ref{SurfaceHamiltonianSM})) and their algebra, where their commutation or anticommutation $XX'-s X'X=0$ are shown with $s=\pm 1$, respectively. 
For the cases where they do not either commute or anticommuting we leave their blocks empty.  The last two columns show the algebra between symmetry operators (edge representation) and $\tau_z$, $\tau_x$.  }
\label{table:1}
\end{table*}

\section{Edge Hamiltonian and symmetry representation}\label{sec:smallM}

In the main text, we argue that due to the chiral propagating of the edge modes in opposite directions, the edge-Hamiltonian is given by,
\begin{equation}\label{SurfaceHamiltonianSM}
	H_{\text{edge}}(\theta, k_{\parallel})= v_{\theta}   \tau_z s_0  k_{\parallel}
\end{equation}
We can justify this edge Hamiltonian as follows.
First $s_z$ is absent from Eq.~(\ref{BdGnewrep2}).
Thus, the edge Hamiltonian is independent of  $s_z$.
Second, we knew that in Eq.~(\ref{BdGnewrep2}) if we change  the sign of $\Delta_s$ (when $\mu=M=0$)  the propagation of chiral modes flips and therefore the edge Hamiltonian has to proportional to $\tau_z$.

In the following, we obtain a representation of the symmetry operators for the edge Hamiltonian.
To derive a representation of symmetry operators we first note that $\mathcal{L}^{\text{edge}}=\tau_z$, and $\mathcal{S}^{\text{edge}}=s_z$. 
Next, we determine the particle-hole operator, regarding the edge-Hamiltonian (anticommuting with Hamiltonian and flip the sign of $k_{\parallel}$), and its relation with $\mathcal{L}^{\text{edge}}$ and $\mathcal{S}^{\text{edge}}$ (given in Table.~\ref{table:1}).
However, the edge Hamiltonian and $\mathcal{L}^{\text{edge}}$ and $\mathcal{S}^{\text{edge}}$ are invariant under $\exp(i\phi+i\alpha \tau_z+i\beta s_z+i\gamma \tau_zs_z)$.
Therefore we have the freedom to choose particle-hole symmetry representation. 
After fixing the particle-hole symmetry representation, we can uniquely determine a representation for  time-reversal symmetry by considering its definition regarding the edge Hamiltonian (commuting with Hamiltonian and flipping the sign of $k_{\parallel}$), and its relation with $\mathcal{S}^{\text{edge}}$, $\mathcal{L}^{\text{edge}}$, and $\Xi^{\text{edge}}$. 
We choose $\Xi^{\text{edge}}=s_x\tau_z\mathcal{K} [k_{\parallel}\rightarrow -k_{\parallel} ]$, and obtain $\Theta^{\text{edge}}=is_y\tau_x\mathcal{K} [k_{\parallel}\rightarrow -k_{\parallel} ]$.

Next, we determine the representation of other spatial symmetries.
Generally, all of the spatial symmetry operators,  have to commute with the edge Hamiltonian. 
To obtain them, we should know how reflection, rotations, and inversion transform $k_{\parallel}$. 
In Table~\ref{table:1} we show the effect of  symmetry operators on $k_{\parallel}$, which can be easily found by considering Fig.~3(a) of the main text (we illustrate rotation and reflection symmetries in Fig.~3(a)). 
If a given symmetry operator flips (keeps)  the sign of $k_{\parallel}$, it has to anticommuting (commute) with $\tau_z$ (we have tabulated them in Table~\ref{table:1}).
For continuing, we  need to find the algebra or relation among different  symmetry operators from their bulk definition (given in Table~\ref{table:1}) and then construct the edge representation for them  by considering the same algebra. 
We  start with $\mathcal{M}^{\text{edge}}_y$. 
Considering definition of $\mathcal{M}^{\text{edge}}_y$ regarding the edge Hamiltonian, and its relations with $\mathcal{S}^{\text{edge}}$, $\mathcal{L}^{\text{edge}}$,  $\Xi^{\text{edge}}$, and $\Theta^{\text{edge}}$, we obtain arbitrary combination of two option. 
This is because  still $H_{\text{edge}}$, $\mathcal{L}^{\text{edge}}$, $\mathcal{S}^{\text{edge}}$, $\Xi^{\text{edge}}$, and $\Theta^{\text{edge}}$ are invariant under $\exp(i\beta s_z)$.
We choose $\mathcal{M}^{\text{edge}}_y=s_y\tau_x [k_{\parallel}\rightarrow -k_{\parallel} ]$ and the representation of other symmetry operators can be determined  uniquely by doing the same procedure (all given in Table~\ref{table:1}). 

As we have discussed in the main letter, the only edge mass that respects both particle-hole and time-reversal operators  is $m(\theta) \tau_x$. 
If we assume that  in the bulk, a $\mathcal{L}$-symmetry-breaking term commute with certain symmetry operations, then similarly its corresponding mass $m(\theta) \tau_x$ has to commute  with the edge representation of those symmetry operators.  
Therefore if the representation of that symmetry operator anticommuting (commute) with $\tau_x$, then $m(\theta)$ should flip (keeps) its sign on the symmetric partner $m(\theta_{\text{s.p}})=-m(\theta)$ [$m(\theta_{\text{s.p}})=m(\theta)$]. 
We also tabulate the commutation or anticommutation of edge representation of  symmetry operators with $\tau_x$ in Table~\ref{table:1}.

Furthermore, under time-reversal and particle-hole symmetry,  we can also add $m'\tau_zs_z $ term to the  edge Hamiltonian Eq.~(\ref{SurfaceHamiltonianSM}).
Considering Eq.~(\ref{SurfaceHamiltonianSM}) and edge mass and $m'\tau_zs_z $ we can write 
\begin{equation}\label{newSurfaceHamiltonianSM}
	H'_{\text{edge}}(\theta, k_{\parallel})= v_{\theta}   \tau_z s_0  k_{\parallel}+m \tau_x+m'\tau_zs_z 
\end{equation}
The energy spectrum of Eq.~(\ref{newSurfaceHamiltonianSM}) is given by
\begin{equation}
    E(k)=\pm \sqrt{(k\pm m')^2+m^2}
\end{equation}
The topological phase transition occurs when $k=\pm m'$ and $m=0$. Nonzero values of $m'$ alter the BOTS transition in momentum space. It is noteworthy that, in the presence of $\lambda_{\text{SO}}$, the BOTS transition does not occur at the time-reversal invariant point (i.e., $k=0$ in our low energy Dirac theory), as demonstrated in Section S6. This observation suggests that spin-orbit coupling introduces additional terms in the effective edge Hamiltonian.

\section{Edge Theory}
We can find the zero energy modes of Eq.~(\ref{BdGnewrep2}) by expanding the Hamiltonian up to the linear order of $k_x$ around $M_1$~\cite{Winkler_Deshpande_2017_PRB_Effective_Hamiltonian_Grphene_M}. 
To describe the zero energy edge modes, we substitute $k_x\to -i\partial_x$, where $x=0$  is the position of the edge which separates interior region $x<0$ from its exterior $x>0$, achieving
\begin{align}\label{Hamilreal}
		\left(-t_1 \zeta(x)	\tfrac{\sigma_x+\sqrt{3}\sigma_y}{2} 
		-i \partial_x (\sqrt{3}\sigma_x -\sigma_y)t_1   \right)\psi_{s_z=\pm1,\tau_z=\pm1}(x)=0
\end{align}
where, we introduce $\zeta(x)=-\text{sgn}(x)$, $|x|\gg0$.
The following wave functions satisfy this equation
\begin{align}\label{psi}
		\psi_{s_z=\pm1,\tau_z=\pm1}(x) {=}\tfrac{1}{\mathcal{N}}e^{\tfrac{1}{2}\int \zeta(x)  dx}  \ket{s_z=\pm1,\tau_z=\pm1,\sigma_z=1}
\end{align}
where $\mathcal{N}$ is the normalization factor.
By projecting $\mu$, $M$, and $t_2$  onto $\psi_{s_z=\pm1,\tau_z=\pm1}(x)$ subspace , we obtain $m_{e4}\equiv m(\theta=0)=-\mu-2 t_2+M$.

\section{Large Sublattice Potential Limit: Effective Hamiltonian}
In the large sublattice potential limit, we can decompose the Hamiltonian of the honeycomb lattice  into two triangular lattices.
For simplicity, consider the Hamiltonian of a honeycomb lattice (normal state) with sublattice potential and nearest-neighbor hopping terms,
\begin{eqnarray}\label{psiH11}
    t_{ij}\psi_j^A-M\psi_i^B=E\psi_i^B,\\
    t_{ij}\psi_j^B+M\psi_i^A=E\psi_i^A,\label{psiH22}
\end{eqnarray}
where $\psi_i^{A}$, and $\psi_i^{B}$ are wave function at site $i$ and sublattice $A$, and $B$ respectively. 
We want to rewrite  Eq.~(\ref{psiH11})-(\ref{psiH22}) in a way that A and B sublattice sectors are decomposed.
To do this, we can rewrite Eq.~(\ref{psiH11})-(\ref{psiH22}) as
\begin{eqnarray}
    \psi_i^B=\tfrac{1}{E+M}t_{ij}\psi_j^A,\\
    \psi_i^A=\tfrac{1}{E-M}t_{ij}\psi_j^B,
\end{eqnarray}
which lead to 
\begin{eqnarray}
   t_{ij}\tfrac{1}{E-M}t_{jk}\psi_k^B-M\psi_i^B=E\psi_i^B \label{eq:eq10},\\
    t_{ij}\tfrac{1}{E+M}t_{jk}\psi_k^A+M\psi_i^A=E\psi_i^A\label{eq:eq20}.
\end{eqnarray}
In the large sublattice limit, we can approximate $E=\mp M$ in the denominator of  Eq.~\ref{eq:eq10}, and Eq.~\ref{eq:eq20}, respectively,
\begin{eqnarray}
   t_{ij}\tfrac{1}{-2M}t_{jk}\psi_k^B-M\psi_i^B=E\psi_i^B \label{eq:eq1},\\
    t_{ij}\tfrac{1}{+2M}t_{jk}\psi_k^A+M\psi_i^A=E\psi_i^A\label{eq:eq2}.
\end{eqnarray}
These Hamiltonians describe two triangular lattices with effective onsite and nearest-neighbor hopping. 
We can read the  effective hopping ($k\ne i$) and onsite potential ($k=i$) as $t^2/(\kappa 2M)$, and $z t^2/(\kappa 2M)+\kappa M$, respectively, where $z$ is the coordination number.

\section{winding number as a topological invariant for effective Kitaev chain}
The topological phase of the effective Kitaev chain also is protected by the chiral symmetry, when $\lambda_R=0$. 
To see this we rewrite the effective Kitaev Hamiltonian  in the basis where both chiral symmetry $\Pi=s_y\tau_y$ and $\mathcal{S}=s_z\tau_z$ are diagonal 
($\ket{\Pi=1,\mathcal{S}=1}$, $\ket{\Pi=1,\mathcal{S}=-1}$, $\ket{\Pi=-1,\mathcal{S}=1}$, $\ket{\Pi=-1,\mathcal{S}=-1}$),
getting a block off-diagonalized matrix
\begin{equation}\label{key}
	U^\dagger H_{\text{Kitaev}}(k) U=\left(\begin{matrix}
		0& h(k)\\ h^{\dagger}(k)&0
	\end{matrix}\right),
\end{equation}
where $U$ is a unitary matrix and
\begin{widetext}
\begin{equation}
h(k)=\left(\begin{matrix}
		-H_N(k)- H_{\text{SO}}(k)+ i H_{\Delta}(k)&  H_{\text{R}}(k)\\
		H_{\text{R}}(k)&-H_N(k)+ H_{\text{SO}}(k)- i H_{\Delta}(k)
	\end{matrix}\right).
\end{equation}
\end{widetext}
When $\lambda_R=0$, the topological invariant of the effective Kitaev chains is given by the winding number of each diagonal element of $h(k)$, or equivalently $\nu_{\mathcal{S}=\pm1}=\tfrac{1}{2\pi}\int dk h^{-1}_{\mathcal{S}\mathcal{S}}(k) \partial_k h_{\mathcal{S}\mathcal{S}}(k)$.
In the topological phase,  the two winding numbers are nonzero but have opposite signs $\nu_{1}=-\nu_{2}=\pm 1$. After turning on $\lambda_R\ne0$, we have to calculate the total winding number of whole $h(k)$ or equivalently $\nu_t=\tfrac{1}{2\pi}\int dk \tr(h^{-1}(k)\partial_k h(k))$, which already gives zero for $\lambda_R=0$. Therefore in the presence of $\lambda_R$, winding number cannot diagnosis the topological phase.

\section{Spin polarized case}
In the Dirac half-metal material, one spin component is gapped out, and the remaining Dirac cones are gapless and spin-polarized. 
Introducing sublattice potential gapped out the remaining Dirac cone.
Alternatively, we can also generalize this material to gapped  Dirac half metal material, where Dirac cones in both spin sectors are gapped out with a spin-dependent sublattice potential $M_{s}\sigma_z$, but sublattice potential for one spin sector is bigger than the other, for instance, $M_{\uparrow}\gg M_{\downarrow}$. 
Therefore, the Fermi contours are polarized in one spin sector, and triplet pairing is the only possible pairing instability.
As we  have discussed before, the existence of normal states (edge modes in the absence of superconductivity) is the key to realizing the BOTS (when superconductivity is turned on), and we can write down an effective Hamiltonian for the spin-polarized Kitaev chain
\begin{eqnarray}\label{polKitaev}
	H_{\text{S.P.K.C}}(k)=H_{\text{N}}(k)\tau_z+H_{\text{SO}}(k)\tau_0 +H_{\Delta}(k)  \tau_y
\end{eqnarray}
In Eq.~(\ref{polKitaev}) $H_{\text{N}}(k)=-\mu+\kappa M_{\downarrow}+2t_2\cos(k)+F^{\kappa}_{\text{boundary}}(k)$, $H_{\text{SO}}(k)=2\kappa\lambda_{\text{SO}}\sin(k)$, $H_{\text{R}}(k)=2\lambda_{\text{R}}\sin(k)$, $H_{\Delta}(k)=2\Delta_{\kappa}\sin(k)$ and $F_{\text{boundary}}(k)=\sum_{n=0}^{n_c} t_{n}^{\prime\kappa}\cos(n k)$ where $t_{n}^{\prime\kappa}$ are the effective hopping terms mediated by other (e.g. bulk) vertices.
The particle-hole and time-reversal symmetry  are  given by $\Xi=\tau_x\mathcal{K} [k\rightarrow -k]$, and $\Theta=\mathcal{K} [k\rightarrow -k]$, respectively.
In the absence of spin orbit coupling, Eq.~(\ref{polKitaev}) is invariant under a chiral symmetry $\tau_x$, and $\mathcal{C}_{2x}=\tau_z [k\to-k]$ symmetry. 
The chiral symmetry allows us to define the winding number and similar to the spin-full calculation, it gives the nontrivial topological phase if $|M|<|\mu+2t|$. 
On the other hand, $Z_2$ topological invariant can be determined by parity information on $k=0,\pi$, which gives similar results to the winding number calculation. 
In the presence of SOC, we cannot use the winding number, but the $Z_2$ topological invariant is well-defined and cannot be changed until the edge Hamiltonian goes to gap-closing.

\section{f-wave spin-triplet pairing and quantum spin hall effect, large spin-orbit coupling limit}

\begin{table*}[ht!]
\centering
\begin{tabular}{|c| c|c| c| c c| c c c c c c c c c c||c|c|c|c|c|} 
\hline
&\multicolumn{2}{|c|}{bulk representation} &\multicolumn{13}{|c|}{algebra between symmetry operators: $XX'- s X'X=0$} & \multicolumn{3}{|c|}{edge representation}& \multicolumn{2}{|c|}{algebra} \\
 \hline
 $X$ & operation & $\mathbf{k}\rightarrow$ &$X^2$& $\Xi$ & $\Theta$&$\mathcal{T}$&$\Tilde{\mathcal{S}}$&$\mathcal{M}_y$&$\mathcal{C}_{2x}$&$\mathcal{P}$&$\mathcal{M}_x$&$\mathcal{C}_{2z}$&$\mathcal{C}_{2y}$&$\mathcal{C}_{3z}$&$\mathcal{C}_{6z}$
 & operation & $\theta\rightarrow$& $k_{\parallel}\rightarrow$&$s_z$& $s_x\tau_x$ \\
 \hline\hline
$\Xi$ & $-i  \tau_x \sigma_z  \mathcal{K}$ & $-\mathbf{k}$
&$+1$& $+1$ & $+1$&$-1$&$+1$&$-1$&$-1$&$-1$&$+1$&$+1$&$+1$&$+1$&$-1$&$\tau_x\mathcal{K} $&$\theta$&$-k_{\parallel}$&$+1$& $+1$ \\
$\Theta$ &$ s_y \mathcal{K} $&$-\mathbf{k}$ 
&$-1$& $+1$ & $+1$&$+1$&$-1$&$-1$&$-1$&$+1$&$-1$&$-1$&$-1$&$+1$&$+1$&$is_y\mathcal{K} $&$\theta$&$-k_{\parallel}$&$-1$& $-1$ \\
$\mathcal{T}$ & $\tau_z$&$\mathbf{k}$  
&$+1$& $-1$ & $+1$&$+1$&$+1$&$+1$&$+1$&$+1$&$+1$&$+1$&$+1$&$+1$&$+1$&$\tau_z$&$\theta$&$k_{\parallel}$&$+1$& $-1$ \\
$\Tilde{\mathcal{S}}$ & $s_z$&$\mathbf{k}$ 
&$+1$& $+1$ & $-1$&$+1$&$+1$&$-1$&$-1$&$+1$&$-1$&$+1$&$-1$&$+1$&$+1$&$s_z$&$\theta$&$k_{\parallel}$&$+1$& $-1$ \\
$\mathcal{M}_y$ & $s_y$&$(k_x,-k_y) $  &
$+1$& $-1$ & $-1$&$+1$&$-1$&$+1$&$-1$&$+1$&$-1$&$-1$&$+1$&$\dots$&$\dots$&$s_y$&$-\theta$&$-k_{\parallel}$&$-1$& $-1$ \\
$\mathcal{C}_{2x}$ & $s_x\tau_z $&$ (k_x,-k_y)$  &
$+1$& $-1$ & $-1$&$+1$&$-1$&$-1$&$+1$&$+1$&$+1$&$-1$&$-1$&$\dots$&$\dots$&$s_x\tau_z$&$-\theta$&$-k_{\parallel}$&$-1$& $+1$ \\
$\mathcal{P}$ & $ \sigma_x $&$- \mathbf{k} $ 
&$+1$& $-1$ & $+1$&$+1$&$+1$&$+1$&$+1$&$+1$&$+1$&$+1$&$+1$&$+1$&$+1$&$ \tau_z$&$\theta+\pi$&$k_{\parallel}$&$+1$& $-1$ \\
$\mathcal{M}_x$ & $s_x\tau_z \sigma_x $&$(-k_x,k_y) $  &$+1$& $+1$ & $-1$&$+1$&$-1$&$-1$&$+1$&$+1$&$+1$&$-1$&$-1$&$\dots$&$\dots$&$s_x$&$\pi-\theta$&$-k_{\parallel}$&$-1$& $+1$ \\
$\mathcal{C}_{2z}$ & $s_z\tau_z \sigma_x$&$- \mathbf{k} $  &$+1$& $+1$ & $-1$&$+1$&$+1$&$-1$&$-1$&$+1$&$-1$&$+1$&$-1$&$+1$&$+1$&$s_z$&$\theta+\pi$&$k_{\parallel}$&$+1$& $-1$  \\
$\mathcal{C}_{2y}$ & $s_y \sigma_x $&$ (-k_x,k_y) $  &$+1$& $+1$ & $-1$&$+1$&$-1$&$+1$&$-1$&$+1$&$-1$&$-1$&$+1$&$\dots$&$\dots$&$s_y\tau_z$&$\pi-\theta$&$-k_{\parallel}$&$-1$& $+1$ \\
$\mathcal{C}_{3z}$& $\tfrac{\mathbbm{1}+i \sqrt{3}s_z\tau_z}{2}$&$  \mathcal{R}_{2\pi/3}\mathbf{k} $  &$-\mathcal{C}^{-1}_{3z}$& $+1$ & $+1$&$+1$&$+1$&$\dots$&$\dots$&$+1$&$\dots$&$+1$&$\dots$&$+1$&$+1$&$\tfrac{\mathbbm{1}+i \sqrt{3}s_z\tau_z}{2}$&$\theta+2\pi/3$&$k_{\parallel}$&$+1$& $+1$ \\
$\mathcal{C}_{6z}$ & $\tfrac{\sqrt{3}\sigma_x+i s_z\sigma_x\tau_z }{2}$&$ \mathcal{R}_{2\pi/6}\mathbf{k} $  &$\mathcal{C}_{3z}$& $-1$ & $+1$&$+1$&$+1$&$\dots$&$\dots$&$+1$&$\dots$&$+1$&$\dots$&$+1$&$+1$&$\tfrac{\sqrt{3}\tau_z+i s_z }{2}$&$\theta+2\pi/6$&$k_{\parallel}$&$+1$& $-1$ \\
 \hline
\end{tabular}
\caption{Symmetry operators (in the bulk representation Eq.~(\ref{lambdaeq2}) and edge representation Eq.~(\ref{SurfaceHamiltonianSMlambda})) and their algebra, where their commutation or anticommutation $XX'-s X'X=0$ are shown with $s=\pm 1$, respectively. 
For the cases where they do not either commute or anticommuting we leave their blocks empty. 
The last two columns show the algebra between symmetry operators (edge representation) and $s_z$, $s_x\tau_x$.  }
\label{table:2}
\end{table*}

In the large $\lambda_{\text{SO}}> M$ limit, the normal state of the system has a nontrivial topological phase, which is known as the quantum spin hall effect~\cite{Kane_Mele_2005_Z2QSHE}.   
In the following,  we derive the edge Hamiltonian using symmetry analysis.
When ($\mu=t_2=M=\lambda_{R}=\Delta_s=\Delta_a=0$), Eq.~(\ref{BdGnewrep}) is reduced to
\begin{equation}\label{lambdaHam}
   	H_{\text{BdG}}(\mathbf{k})=t_1 H^{\text{++}}_1(\mathbf{k})\sigma_x\tau_z+t_1 H^{\text{-+}}_{1'}(\mathbf{k})\sigma_y\tau_z+ \lambda_{SO}  H^{\text{+-}}_3(\mathbf{k})\mathbf{s}_z\sigma_z 
\end{equation}
This Hamiltonian is invariant under  
 $\Xi$, $\Theta$, $\mathcal{C}_{6z}$, $\mathcal{C}_{2z}$,  $\mathcal{C}_{2y}$, $\mathcal{M}_x$,  $\mathcal{P}$, $\mathcal{C}_{3z}$, $\mathcal{C}_{2x}$, $\mathcal{M}_y$, $\mathcal{S}$, $\mathcal{T}$, and $\Tilde{\mathcal{S}}$. In this section for simplicity, we assume $\lambda_R=0$.
 
It is helpful to rewrite Eq.~(\ref{lambdaHam}) in a new basis using a unitary transformation 
\begin{equation}
    U=\exp^{\tfrac{1}{4}i\pi (\sigma_z-\sigma_z\tau_z)},
\end{equation}
which leads to
\begin{align}\label{lambdaeq2}
    U^\dagger H_{\text{BdG}}(\mathbf{k})U=
 t_1 H^{\text{++}}_1(\mathbf{k})\sigma_x+t_1 H^{\text{-+}}_{1'}(\mathbf{k})\sigma_y+\lambda_{\text{SO}} H^{\text{+-}}_3(k)\sigma_z s_z.
\end{align}
The Eq.~(\ref{lambdaeq2}) describes two copies ( $\tau_z=\pm 1$) of the celebrated Kane-Mele model, and hosts two chiral boundary modes propagating in opposite directions.
Therefore the edge Hamiltonian can be described by
\begin{equation}\label{SurfaceHamiltonianSMlambda}
	H^0_{\text{edge}}(\theta, k_{\parallel})= v_{\theta}  k_{\parallel} s_z \tau_0.
\end{equation}
We obtain the symmetries representation of Eq.~(\ref{SurfaceHamiltonianSMlambda}), using the same procedure that we have done in Sec.~\ref{sec:smallM}, where we tabulate them in Table.~\ref{table:2}. 

In the following, we try to find a general edge Hamiltonian regarding nonzero $\Delta_s$, $\mu$, $M$, and  $\Delta_a$.
Each of these terms has a certain commutation or anticommutation relation with symmetry operators and
we assume that the main effect of them also respects those relations in the edges Hamiltonian. 

\begin{figure}
 	\centering
 	\includegraphics[width=\linewidth]{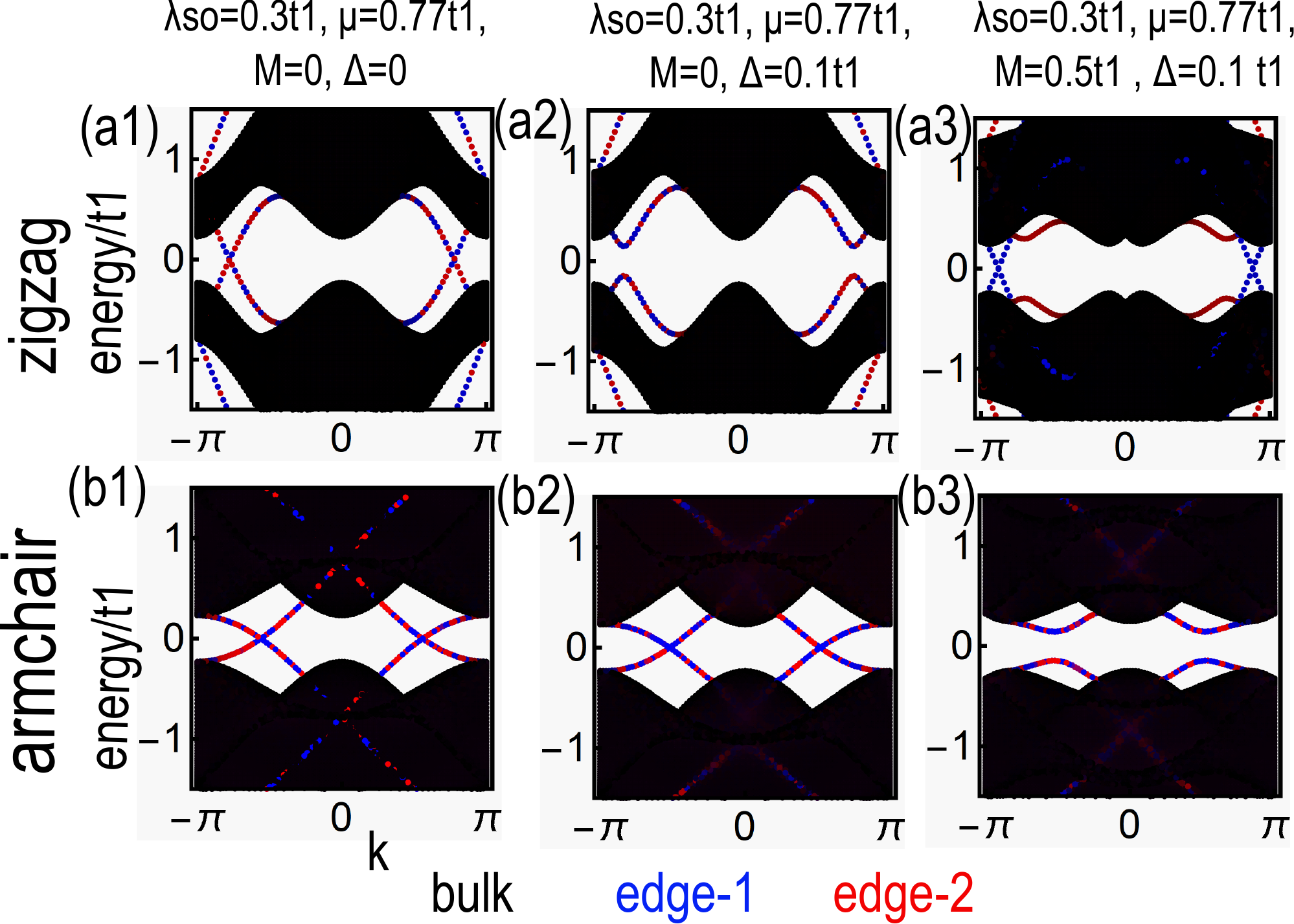}
 	\caption{
 			Energy dispersion (superconducting states, or BdG Hamiltonian) of a strip geometry of the honeycomb lattice   with f-wave spin-triplet pairing and spin-orbit coupling.  }
 	\label{fig:soc}
 \end{figure}

The $\mu$, $t_2$ and $M$ respect all of the local symmetries ($\Xi$, $\Theta$, $\mathcal{T}$, $\Tilde{\mathcal{S}}$, and $\mathcal{S}$). 
The only $k_{\parallel}$ independent term that can satisfy these symmetries is 
\begin{equation}
   W_{\alpha}(\theta) \tau_z, 
\end{equation}
where $\alpha=\mu, t_2, M$. 
These  terms, are symmetric under $\mathcal{M}_{y}$, $\mathcal{C}_{2x}$ and $\mathcal{C}_{3z}$.
Accordingly and using Table.~\ref{table:2}, $W_{\alpha}(\theta)=W_{\alpha}(-\theta)=W_{\alpha}(\theta+2\pi/3)$.
However, although $\mu$, and $t_2$ are symmetric  under $\mathcal{M}_{x}$, $\mathcal{C}_{2y}$, $\mathcal{P}$, $\mathcal{C}_{2z}$ and $\mathcal{C}_{6z}$ (commute with them), but $M$ not (anticommute with them). 
Thus, $\mu$ and $t_2$ leads to (where $\alpha=\mu, t_2$)
\begin{equation}
    W_{\alpha}(\theta)=W_{\alpha}(\pi-\theta)=W_{\alpha}(\pi+\theta)=W_{\alpha}(\theta+2\pi/6),
\end{equation}
while $M$ leads to
\begin{equation}
    W_{M}(\theta)=-W_{M}(\pi-\theta)=-W_{M}(\pi+\theta)=-W_{M}(\theta+2\pi/6). 
\end{equation}
This means that in the $\mathcal{M}_x$  symmetric edges (armchair edges) $W_{M}(\theta)=0$.
Note that $W_{\mu, t_2, M}(\theta) \tau_z$, does not commute with Hamiltonian and cannot gap out the edge Hamiltonian (see Fig.~\ref{fig:soc}(a1), and Fig.~\ref{fig:soc}(b1)). 

As we discussed before, nonzero $\Delta_s\ne0$ (or $\Delta_a\ne0$) breaks $\mathcal{T}$, and $\Tilde{\mathcal{S}}$, but still system is invariant under $\mathcal{S}$. 
If we assume three remaining local symmetries $\Xi$, $\Theta$, and $\mathcal{S}$, we can add the following terms to the edge Hamiltonian, that break $\Tilde{\mathcal{S}}$, and $\mathcal{T}$
\begin{equation}
    \delta_{\alpha}(\theta) s_y\tau_y+\Delta_{\alpha}^{\text{edge}} (\theta) k_{\parallel}s_x\tau_x,\qquad \alpha=\Delta_s, \Delta_a.
\end{equation}
These terms anticommuting with Eq.~(\ref{SurfaceHamiltonianSMlambda}) and can gap it out (Fig.~\ref{fig:soc}(a2)).
Note that even with nonzero  $\Delta_s$  or $\Delta_a$, the system is invariant under $\mathcal{M}_{y}$, $\mathcal{C}_{2x}$ and $\mathcal{C}_{3z}$. 
Using Table.~\ref{table:2} we obtain
\begin{equation}\delta_{\alpha}(\theta)=\delta_{\alpha}(-\theta)=\delta_{\alpha}(\theta+2\pi/3),\end{equation} and 
\begin{equation}\Delta_{\alpha}^{\text{edge}}(\theta)=\Delta_{\alpha}^{\text{edge}}(-\theta)=\Delta_{\alpha}^{\text{edge}}(\theta+2\pi/3).\end{equation} 
Although $\Delta_s$ is symmetric (commute with them) under $\mathcal{M}_{x}$, $\mathcal{C}_{2y}$, $\mathcal{P}$, $\mathcal{C}_{2z}$ and $\mathcal{C}_{6z}$, but $\Delta_a$ are not (anticommute with them). 
Using Table.~\ref{table:2}, we obtain 
\begin{equation}\delta_{\Delta_s}(\theta)=-\delta_{\Delta_s}(\pi+\theta)=-\delta_{\Delta_s}(\pi-\theta),\end{equation}
\begin{equation}\Delta^{\text{edge}}_{\Delta_s}(\theta)=-\Delta^{\text{edge}}_{\Delta_s}(\pi+\theta)=-\Delta^{\text{edge}}_{\Delta_s}(\pi-\theta),\end{equation}
\begin{equation}\delta_{\Delta_a}(\theta)=\delta_{\Delta_a}(\pi+\theta)=\delta_{\Delta_a}(\pi-\theta),\end{equation} and
\begin{equation}\Delta^{\text{edge}}_{\Delta_a}(\theta)=\Delta^{\text{edge}}_{\Delta_a}(\pi+\theta)=\Delta^{\text{edge}}_{\Delta_a}(\pi-\theta).\end{equation}
This means that at the $\mathcal{M}_x$  symmetric edges (armchair edges) $\delta_{\Delta_s}(\theta)=\Delta^{\text{edge}}_{\Delta_s}(\theta)=0$.

We can write general edge Hamiltonian as follows
\begin{equation}\label{SurfaceHamiltonianSMlambda3}
	H_{\text{edge}}(\theta, k_{\parallel})= v_{\theta}  k_{\parallel} s_z+W(\theta) \tau_z+ \delta(\theta) s_y\tau_y+\Delta^{\text{edge}} (\theta) k_{\parallel}s_x\tau_x,
\end{equation}
where 
\begin{align}
&W(\theta)=W_{\mu}(\theta) +W_{t_2}(\theta) +W_{M}(\theta)+W_{\text{other corrections}}(\theta) ,\\
&\delta(\theta)=\delta_{\Delta_s}(\theta) +\delta_{\Delta_a}(\theta) +\delta_{\text{other corrections}}(\theta), \\
&\Delta^{\text{edge}} (\theta)=\Delta^{\text{edge}} _{\Delta_s}(\theta) +\Delta^{\text{edge}} _{\Delta_a}(\theta) +\Delta^{\text{edge}} _{\text{other corrections}}(\theta).
\end{align}

As we have discussed before,  $\mathcal{M}_{y}$, $\mathcal{C}_{2x}$ and $\mathcal{C}_{3z}$ gives
\begin{align}
 &W(\theta)=W(-\theta) =W(\theta+2\pi/3), \\
  &\delta(\theta)=\delta(-\theta) =\delta(\theta+2\pi/3), \\
&\Delta^{\text{edge}} (\theta)=\Delta^{\text{edge}} (-\theta) =\Delta^{\text{edge}} (\theta+2\pi/3).
   \end{align}
   However, when $M=\Delta_a=0$ then $\mathcal{M}_{x}$, $\mathcal{C}_{2y}$, $\mathcal{C}_{2z}$, $\mathcal{P}$ and $\mathcal{C}_{6z}$ gives
 \begin{align}
 &W(\theta)=W(\pi-\theta) = W(\pi+\theta)=W(\theta+2\pi/6), \\
&\delta(\theta)=-\delta(\pi-\theta) = -\delta(\pi+\theta)=-\delta(\theta+2\pi/6), \\
&\Delta^{\text{edge}}(\theta){=}-\Delta^{\text{edge}}(\pi{-}\theta){=}-\Delta^{\text{edge}}(\pi{+}\theta)=-\Delta^{\text{edge}} (\theta{+}\frac{2\pi}{6}). 
   \end{align}

In the following, we discuss different topological phases of the system by studying the gapless state of the general edge Hamiltonian Eq.~(\ref{SurfaceHamiltonianSMlambda3}) that is forced by symmetries relations. 
We first discuss the case of $\mu=t_2=M=\Delta_a=0$, then we turn on them and discuss possible topological phases.

When $\mu=t_2=M=0$ (but nonzero $\Delta_s$ and $\Delta_a$), the bulk Hamiltonian is invariant under two other local symmetries $\mathcal{O}_1=\tau_x\sigma_zs_z$, and $\mathcal{O}_2=\tau_y\sigma_z$ (in the representation of Eq.~(\ref{BdGnewrep})).
We obtain their edge Hamiltonian representation as $\mathcal{O}^{\text{edge}}_1=\tau_x$ and $\mathcal{O}^{\text{edge}}_2=s_z\tau_y$, by using their relationships (commutation or anticommutation) with other local symmetries $\Xi$, $\Theta$, etc.
Note that $\mathcal{O}^{\text{edge}}_1$ and $\mathcal{O}^{\text{edge}}_2$ force $W(\theta)=\delta(\theta)=0$ in Eq.~(\ref{SurfaceHamiltonianSMlambda3}).
Consequently, Eq.~(\ref{SurfaceHamiltonianSMlambda3}) remains gapless and f-wave pairing potential can not gap out the normal state, and the system can be considered as a first-order topological phase, which indicates gapless states exist along all boundaries. This phase is connected  ($\Delta_a=0$, $\lambda_R\rightarrow 0$)  to the first-order topological state that we discuss in the main letter.

If we turn on $\mu\ne 0$ or $t_2\ne0$, but keep $M=\Delta_a=0$, a nonzero $W(\theta)$, $\delta(\theta)$, and $\Delta^{\text{edge}}(\theta)$ can be induced in the edge Hamiltonian.
However, symmetry constraints (type-1 symmetry group) force sign changing for $\Delta^{\text{edge}}(\theta_{s.p})=-\Delta(\theta)$, and $\delta(\theta_{s.p})=-\delta^{\text{edge}}(\theta)$, which means that they vanish at the symmetric edges (armchair edges).
Therefore, the normal modes remain gapless, and f-wave spin-triplet pairing does not gap them out  in the $\mathcal{M}_x$ symmetric edges (compare Fig.~\ref{fig:soc}(a2), and Fig.~\ref{fig:soc}(b2)).
We can call this system, a crystalline topological superconductor.
Furthermore, due to the sign changing of $\delta(\theta)$, and $\Delta^{\text{edge}}(\theta)$, this system becomes a HOTS in the absence of symmetric edges\cite{scammell2021intrinsic}.

By introducing $M\ne0$, or $\Delta_a\ne0$,  type-1 symmetries will be broken, and  there is no constraint that force  gapless states for Eq.~(\ref{SurfaceHamiltonianSMlambda3}).
Therefore, the system is not any more HOTS, and crystalline topological superconductivity breaks down.
Consequently, gapless edge modes along $\mathcal{M}_x$ symmetric edges are gaped out (see Fig.~\ref{fig:soc}(b3)). 
However, the system still can obtain a nontrivial BOTS phase, where the topological phase transition from the nontrivial phase to the trivial phase mediated by gap closing at a $\mathcal{M}_y$ symmetric edge. 
The energy spectrum of Eq.~(\ref{SurfaceHamiltonianSMlambda3}) is zero (gapless) if $W(\theta)=-v(\theta)\tfrac{\delta(\theta)}{\Delta^{\text{edge}}(\theta)}$, and $k_{\parallel}=\tfrac{\delta(\theta)}{\Delta^{\text{edge}}(\theta)}$. 
Now consider two $\mathcal{M}_y$ symmetric edges that are located at $\theta=0, \pi$. 
Doing the same procedure that we have done in the main letter,  we can find $W(0)\approx -(\mu+2t_2)+M$, while $W(\pi)\approx -(\mu+2t_2)-M$.
Therefore, in the absence of $\delta(0, \pi)=0$, the BOTS phase transition is given by  $|M|=|\mu+2t_2|$, which is the same as that we obtain in other limits. 
This BOTS phase transition is not dependent on spin-orbit coupling strength, so reducing $\lambda_{so}$ does not change it.
Finally, a trivial insulator is  deriving from the topological quantum spin hall phase by reducing $\lambda_{so}$. 
Though in this process the normal state undergoes a bulk gap closing, in the superconducting phase this process is always gapped.
This phase is thus connected to the large sublattice potential limit that we discussed in the main letter.

In Fig.~\ref{fig:soc}(a3) we plot the BOTS phase transition, which appears for $k\ne0,\pi$ (corresponding to $k=0$ in our low energy Dirac theory),  indicates nonzero $\delta(\theta)$. 
Note that $\delta(\theta)$ breaks $\mathcal{T}$ symmetry and is a superconducting pairing potential. 
Because $ \delta(\theta) s_y\tau_y$ is $k_{\parallel}$ independent, we can interpret it as an effective spin-singlet pairing potential. 
We confirm that in the absence of $\lambda_{so}$ the BOTS transition happens in $k=0$, indicating $\delta(\theta)$ is emerging from a combination of spin-orbit coupling, f-wave spin-triplet pairing, and sublattice potential processes.

\section{f-wave spin-triplet pairing and valley quantum hall effect }

\begin{figure}[t!]
 	\centering
 	\includegraphics[width=\linewidth]{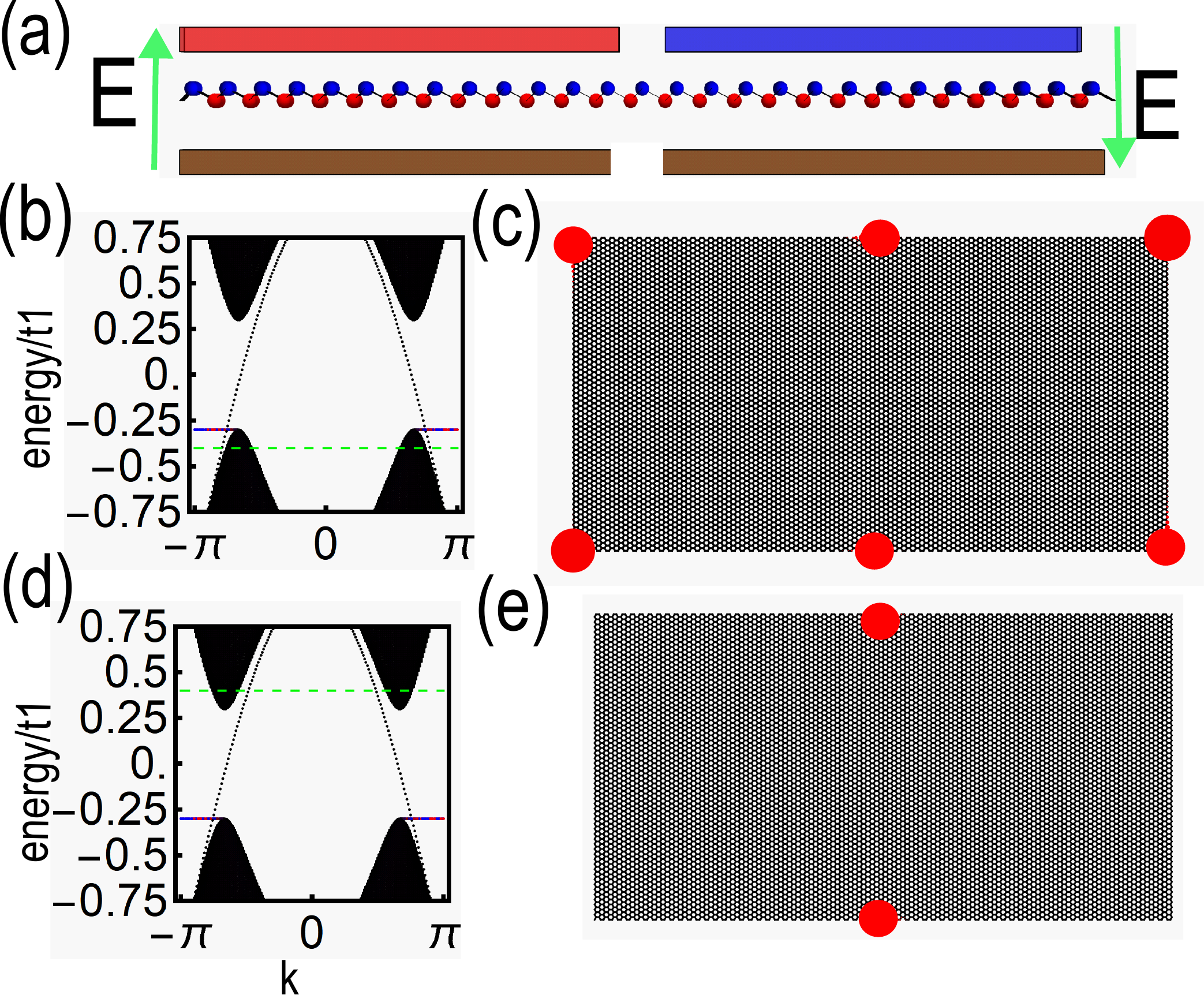}
 	\caption{
  The f-wave spin-triplet pairing and valley quantum hall effect.
  (a) The double gate setup with an opposite electric field on the opposite halves of the system. 
  (b), (d) The energy dispersion (normal state) of a zigzag strip geometry with an opposite sublattice mass  $|M|=0.3 t_1$ ($t_1=1$) on the  two sides of the system. 
(c), (e) Majorana zero-energy modes (red spots), when we turn on f-wave spin-triplet pairing  $\Delta_s=0.1t_1$.
We set (c) $\mu=-0.4$, (e) $\mu=0.4$ which is shown by  green dashed lines in (b), and (d), respectively. 
  The number of Majorana zero modes in (c) and (e) amount to 12, and 4, respectively.
  }
 	\label{fig:valley}
 \end{figure}

Boundary modes in the honeycomb structures can be achieved by applying opposite sublattice potential on the two sides of the system. 
This can be done by applying an opposite electric field on two opposite sides of a buckled honeycomb lattice (Fig.~\ref{fig:valley}(a)), in which a boundary mode appears in the middle of them (Fig.~\ref{fig:valley}(b)). 
Actually, this boundary mode is related to the edge mode of a topological phase with a nonzero valley Chern number (equal to one) called the valley quantum hall effect and  is protected by the absence of scattering between two valleys \cite{Youngkuk_Valley_Silicene_2014}. 
One can imagine this boundary mode as a 1D chain that lives in the middle of the system. 
We show in Fig.~\ref{fig:valley}(c), and Fig.~\ref{fig:valley}(e)  that the Majorana zero-energy modes appear (when $\Delta_s\ne0$) at the ends of this effective 1D (Kitaev) chain. 
The corner modes in Fig.~\ref{fig:valley}(c), are related to the normal modes that appear at the zigzag edges (two effective Kitaev chain that lives on the two sides of the system).
These normal modes  obtain the same energy due to opposite sublattice potential on the two sides of the system (see degenerate blue and red edge states of the energy spectrum in Fig.~\ref{fig:valley}(b)).

\section{topological invariant for the strip of the honeycomb lattice}

\begin{figure}[t!]
	\centering
	\includegraphics[width=\linewidth]{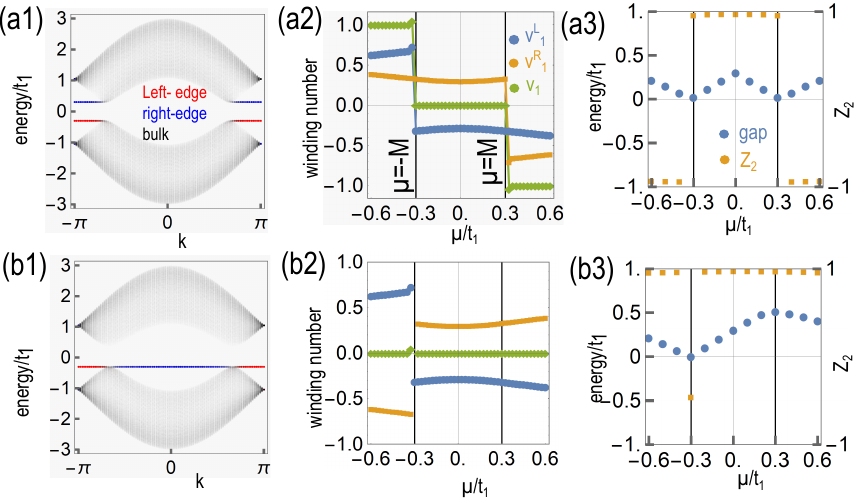}	
	\caption{
		The BOTS transition in a strip geometry where (a) two edges are zigzag,  and (b) one is zigzag and the other is dangling. 
We plot energy dispersion (normal state), the edge resolved winding number, and the $Z_2$ topological invariant in the left, center, and right panels. 
In this calculation, we set $M=0.3t_1$, $\Delta_s=0.1t_1$, and other parameters to zero. 
We also show the energy gap (in the superconducting state) in the right panels.
	}
	\label{fig:Fig5}
\end{figure}

We can generalize the $Z_2$ topological invariant  or winding number to calculate the topological states of the strip geometry. 
To confirm the Kitaev chain picture, we define edge resolved winding-number $\nu_{\mathcal{S}=\pm1}=\nu^{\text{L}}_{\mathcal{S}}+\nu^{\text{R}}_{\mathcal{S}}$, where  $\nu^{L, R}_{\mathcal{S}}=\tfrac{1}{2\pi}\int dk \text{Tr}^{L,R} (\tilde{h}^{-1}_{\mathcal{S}}(k) \partial_k \tilde{h}_{\mathcal{S}}(k))$, $\tilde{h}_{\mathcal{S}}$  is the off-diagonal part of the BdG Hamiltonian in the basis of chiral symmetry and $\mathcal{S}$. 
The $\text{Tr}^{L,R}$ is the partial trace that only sum over indices that belongs to the left (L) or right (R) part of the system (two opposite half of the strip geometry, see Fig.~1(a)) ~\cite{2020BOWuTP_SC_IronPnictides}.  
Note that although $\nu_{\mathcal{S}}$ is a quantize number, but $\nu^{R, L}_{\mathcal{S}}$ are not.
 However, the jumping of $\nu^{R, L}_{\mathcal{S}}$ across the topological phase transition is a good signature for topological phase transition~\cite{2020BOWuTP_SC_IronPnictides}.
Furthermore, we numerically calculate the $Z_2$ topological invariant for the strip geometry (see Ref.~\onlinecite{PhysRevB.88.134523} for the procedure). 
In Fig.~\ref{fig:Fig5}(a1)-(a3) we compare the  $Z_2$, winding number and  energy gap for a system with  zigzag edges, which confirms the applicability of these topological invariants for diagnosing topological phases. 

\section{BOTS at dangling edge}
The fact that the existence of a nontrivial topological  phase with corner modes depends on the normal state condition, leads us to design different edge terminations and therefore manipulate corner modes.
For instance, suppose in designing strip geometry, we instead terminate two edges with the same sublattice atoms (for instance remove the red vertices on the right side of Fig.~1(a)).
In this case, one edge is zigzag and the other is dangling. 
The coordination number on the dangling edge is $z=1$ and we expect normal modes for them.
However, its energy is  the same  as the zigzag normal modes (compare Fig.~\ref{fig:Fig5}(a1), Fig.~\ref{fig:Fig5}(b1)).  
Therefore,  the two edges obtain the same topological state simultaneously, where the jump of edge-resolved winding number  shows topological phase transition for both edges (see Fig.~\ref{fig:Fig5}(b2)).
Note that, the $Z_2$ and  $\nu_{\mathcal{S}}$ cannot diagnose this topological phase transition [see Fig.~\ref{fig:Fig5}(b3)].
Remarkably, because of two nontrivial Kitaev chains on both sides of the system, there is no guarantee that the topological phase transition happen at a symmetric edge. 

\clearpage

 \bibliography{refs}